\documentclass[useAMS,usenatbib,]{mn2e}
\usepackage{times}
\usepackage{natbib}
\usepackage[dvips]{graphicx}
\usepackage{subfigure}
\usepackage{graphicx}
\usepackage{amsmath}
\usepackage{txfonts}
\usepackage{natbib}
\usepackage{threeparttable}
 
\DeclareMathOperator{\atan}{atan}

\title[] {Measurements  and simulation of Faraday rotation across the Coma radio relic}
\author[A. Bonafede et al. ]
{A. Bonafede$^{1}$\thanks{E-mail: annalisa.bonafede@hs.uni-hamburg.de}, F. Vazza$^{1,2}$, M. Br\"uggen$^{1}$, M. Murgia$^{3}$, F. Govoni$^{3}$, L. Feretti$^{2}$,
\newauthor G. Giovannini$^{2,4}$, G. Ogrean$^{1}$ \\
$^1$ Hamburger Sternwarte, Universit\"at Hamburg, Gojenbergsweg 112, 21029, Hamburg, Germany. \\
$^2$ INAF Istituto di Radioastronomia, via P. Gobetti 101, I-40129 Bologna, Italy. \\
$^3$ INAF Osservatorio di Cagliari, Italy. \\
$^4$ Universit\`a di Bologna, Dip. di  Fisica e Astronomia, via Ranzani 1, I-40127 Bologna, Italy. 
}

\begin{document}  
   \date{Received ; accepted }

\maketitle

\begin{abstract}  
   The aim of this work is to probe the magnetic
   field properties in relics and infall regions of galaxy clusters using Faraday Rotation Measures. We present  Very Large Array
   multi-frequency observations  of
      seven sources in the region South-West of the Coma cluster, where
   the infalling group NGC4839 and the relic 1253+275 are located.  The
   Faraday Rotation Measure maps for the observed sources are
   derived and analysed to study the magnetic field in the South-West region of Coma.  We discuss how to
   interpret the data by comparing observed and mock
   rotation measures maps that are produced simulating different
   3-dimensional magnetic field models. The magnetic field model that  gives the best fit to the Coma central region
   underestimates the rotation measure in the South-West region by a
   factor $\sim$6, and no significant jump in the rotation measure data is
   found at the position of the relic.  We explore different possibilities to reconcile 
   observed and mock rotation measure trends, and conclude that an amplification of the magnetic field
   along the South-West sector is the most plausible solution.   
      Our data together with recent X-ray estimates of the gas density 
   obtained with {\it Suzaku} suggest that a magnetic field amplification by a factor $\sim$3 is required throughout the entire
   South-West region in order to reconcile real and mock rotation measures trends.
   The magnetic field in the relic region is inferred to be $\sim 2 \, \mu$G,  consistent with Inverse Compton limits.

 \end{abstract}



\section{Introduction}
\begin{figure*}
\centering
\includegraphics[width=0.6\textwidth]{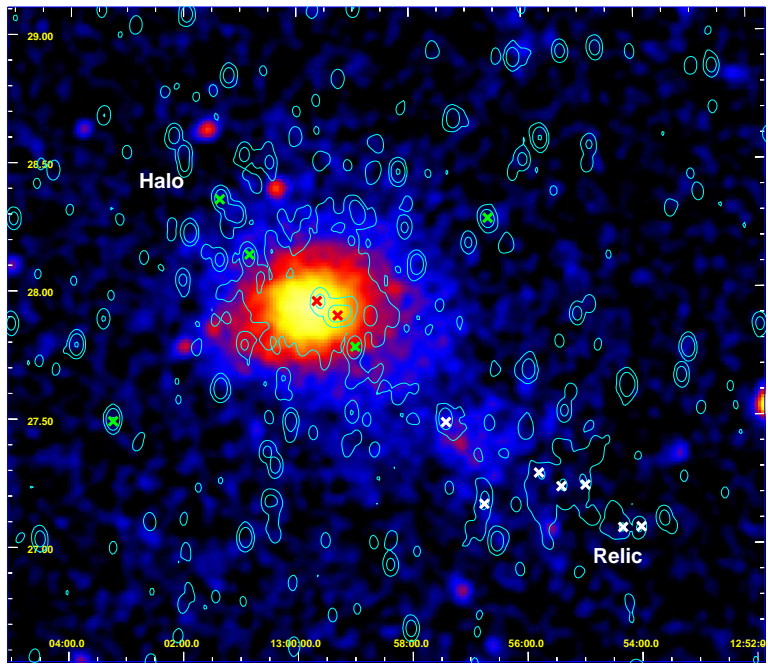}
\caption{Colours: X-ray emission from the Coma cluster and the NGC4839
group from the Rosat All Sky Survey \citep{1992A&A...259L..31B}. Contours: Radio emission from
WSRT  at 325 MHz \citep{Venturi90}. Contours level are at 0.3, 1 and 3 mJy/beam. The
beam is $\sim 50" \times 125"$. Crosses mark the position of the sources in the Coma cluster
(green and red) and in the NGC4839 group (white crosses), analysed in
this work.}
\label{Xradio}
\end{figure*}

The hot and rarefied plasma that fills the intra-cluster medium (ICM)
of galaxy clusters is known to be magnetised.  The existence of the
magnetic field in galaxy clusters is inferred from radio observations of
radio halos and relics, which are synchrotron emitting sources not
obviously connected to any of the cluster radio galaxies. Nowadays,
this emission is detected in $\sim$ 70 objects \citep{Feretti12}.
Radio relics are diffuse extended radio sources in the outer regions
of galaxy clusters. The radio emission indicates the presence of
relativistic particles and magnetic fields \citep[see reviews by][and
  references therein]{2011SSRv..tmp..138B}.  The origin of relics is
still uncertain: there is a general consensus that it is related to
shock waves occurring in the ICM during merging events.  Cosmological
simulations indicate that shock waves are common phenomena during the
process of structure formation, and the Mach numbers of merging shocks are expected to
be of the order of $\sim$ 2-4
\citep[e.g.][]{sk08,2012arXiv1211.3122S,va10kp,2011MNRAS.418..960V,2011ApJ...734...18K,2003ApJ...593..599R}. However,
the direct detection of shock waves is difficult, because of the low
X-ray surface brightness of the clusters at their periphery. So far,
only a handful of shock fronts have been unambiguously detected
through both a temperature and a surface brightness jump: the ``bullet
cluster'' - 1E 065756 - \citep{2002ApJ...567L..27M}, Abell 520
\citep{2005ApJ...627..733M}, Abell 3667
\citep{2010ApJ...715.1143F,2011arXiv1112.3030A}, Abell 2146
\citep{2011MNRAS.417L...1R}, Abell 754 \citep{2011ApJ...728...82M},
CIZAJ2242.8+5301 \citep{2011arXiv1112.3030A,OgreanCiza}, and
Abell 3376 \citep{2011arXiv1112.3030A}. Among these clusters, Abell
3667, CIZAJ2242.8+5301, Abell 754, and Abell 3376 host one or two radio relics,
a radio halo is found in the bullet cluster and
in Abell 520, while no radio emission is detected in Abell
2146. Hence, there is not always a one-to-one connection between a radio relic
and a shock wave detected through X-ray observations. In addition, the
mechanism responsible for the particle acceleration is not fully
understood yet. Shock waves should amplify the magnetic field and
(re)accelerate protons and electrons to relativistic energies,
producing detectable synchrotron emission in the radio domain \citep{1998A&A...332..395E,2012MNRAS.423.2781I}. 
The process responsible for the particle acceleration could be either
direct Diffusive Shock Acceleration or shock re-acceleration of previously
accelerated particles
\citep[e.g.][]{2007MNRAS.375...77H,2011ApJ...728...82M,2012arXiv1205.1895K,2011ApJ...734...18K}.
Regardless of the origin of the emitting relativistic electrons, a key
role is played by the magnetic field. Previous magnetic field
estimates in radio relics have been obtained using three
different approaches:
\begin{itemize}
\item{} radio equipartition arguments, resulting in magnetic field
  estimates of $\sim 1 \; \mu\rm{G}$ \citep[see e.g. review by][]{Feretti12}. \\
\item{} Inverse Compton (IC) studies, resulting mostly in lower limits
  on the magnetic field strength $> 1 \; \mu\rm{G}$ \citep[see][and references therein]{2008MNRAS.383.1259C}.\\
\item{} Width of the relic, as proposed by
  \citet{2007MNRAS.375...77H}- only applied so far by
  \citet{2010Sci...330..347V}- resulting in magnetic field estimates
  of $0.2 - 7 \mu\rm{G}$. This technique assumes that a
  shock wave is present at the location of the observed radio emission.
\end{itemize}
Equipartition estimates rely on several assumptions that are
difficult to verify. IC emission is hardly detected from radio
relics, hence leading mostly to average lower limits for the magnetic
field strength. The method proposed by \citet{2007MNRAS.375...77H} can
be applied only in particular circumstances \citep[see][for
details]{2010Sci...330..347V}.  \\
Another method to analyse the magnetic field is the study of the
Faraday Rotation.  The interaction of the ICM, a magneto-ionic medium,
with the linearly polarized synchrotron emission results in a rotation
of the wave polarization plane (Faraday Rotation), so that the
observed polarization angle, $\Psi_{\rm obs}$ at a wavelength $\lambda$
differs from the intrinsic one, $\Psi_{\rm int}$ according to:
\begin{equation}
\label{eq:psiRM}
 \Psi_{\rm obs}(\lambda) = \Psi_{\rm int}+\lambda^2 \times RM,
\end{equation}
where RM is the Faraday Rotation Measure. This is related to the
magnetic field component along the line-of-sight ($B_{//}$)
weighted by the electron number density ($n_e$) according to:
\begin{equation}
\label{eq:RM}
RM \propto \int_{los}n_e(l) B_{//}(l) dl.
\end{equation}
Statistical RM studies of clusters and background sources show that
magnetic fields are common in both merging and relaxed clusters
\citep{2004JKAS...37..337C,2004mim..proc...13J}, although their origin
and evolution are still uncertain.  Information about the magnetic
field strength, radial profile and power spectrum can be obtained by
analyzing the RM of sources located at different distances from the
cluster center \citep[see
  e.g.][]{Murgia04,2006A&A...460..425G,2008A&A...483..699G,Bonafede10,2010A&A...522A.105G}. These
studies indicate that the magnetic field strength decreases from the
center to the periphery of the cluster in agreement with predictions
from cosmological simulations
\citep[e.g.][]{1999A&A...348..351D,2005ApJ...631L..21B,2009MNRAS.398.1678D,2011ApJ...740...81R,Bonafede11b}.
Observations and simulation indicate that the magnetic field in
cluster outskirts is a fraction of $\mu\rm{G}$.  So far, the RM method
has not been applied to relics and infall regions, but only to the
central regions of a few galaxy clusters. This is partly due to the difficulty of finding
several sources - located in projection through the region that one wants to analyse -
that are both extended and sufficiently strong to derive the RM distribution
with the desired accuracy.\\ In this work,  we
investigate the magnetic field strength and topology in the region of
the relic in the Coma cluster using RM analysis of sources located in
the region of the relic. The relic in the Coma cluster is among the
best candidates for this kind of study: it has a large angular extent
($\sim$ 0.5$^{\circ}$), hence some radio sources are seen in
projection through it; it was the first relic that was discovered,
making it one of the most studied, and often the test-case for
theoretical models \citep[e.g.][]{1998A&A...332..395E}; in addition, recent radio \citep{BrownRudnick11}
and X-ray \citep{OgreanComa,2013arXiv1302.2907A,2013arXiv1302.4140S} observations suggest the presence of a
long filamentary structure South-West of the Coma cluster, associated
with the NGC4839 group and with the relic.  \citet{OgreanComa} and \citet{2013arXiv1302.2907A} found a
possible $M \sim 2$ shock, in a region spatially coincident with the
relic, although such a shock has been detected only in temperature and
not in density or surface brightness. Finally, previous studies of
Coma have already set constraints on the magnetic field strength and
structure of the cluster \citep{Bonafede10}. In \citet{Bonafede10}, we
have studied the Faraday RM of several sources located in the Coma
cluster, excluding the South-West region where the infall group is
located. Through a comparison with mock RM images, we have derived the
magnetic field model which best reproduces the observed RMs within the central 1.5 Mpc region. The best
agreement with data is achieved with a magnetic field central strength of
4.7 $\mu$G, and a radial slope that follows the square-root of gas
density profile. Starting from these results, it is possible to
estimate whether a magnetic field amplification is required in the
relic region. An overview of the sources's position with respect to the Coma cluster 
is displayed in Fig. \ref{Xradio}.\\ The paper is organised as follows: radio observations
and data reduction are reported in Sec. \ref{sec:Obs}, and the results
of the Faraday RM are reported in Sec. \ref{sec:RMobs}. In
Sec. \ref{sec:code} we describe the technique we use to produce mock
RM observations, and in Sec. \ref{sec:simulations} mock and observed RM images are analysed and  compared . 
Our results are
discussed in Sec. \ref{sec:discussion} and finally, we conclude in
Sec. \ref{sec:conclusion}.\\ Throughout this paper,s we assume a
concordance cosmological model $\rm{\Lambda CDM}$, with $H_0=$ 71 km
s$^{-1}$ Mpc$^{-1}$, $\Omega_M=$ 0.27, and $\Omega_{\Lambda}=$
0.73. One arcsec corresponds to 0.46 kpc at $z=$0.023.

\section{Radio observations}
\label{sec:Obs} 
\subsection{VLA observations and data reduction}
From the FIRST survey \citep{FIRST}, we selected the 6 brightest and
extended radio galaxies that appeared in the NVSS to be polarised
\citep{NVSS}. Three of them are well-known Coma radio galaxies:
5C4.51, the central galaxy of the merging group referred by us as the NGC 4839 group, 5C4.43, and
5C4.20. The other radio-sources are  background objects.  In
the field of view of 5C4.20 another source was found to be unresolved
and polarised: 5C4.16, which has been included in our analysis.
Observations were carried out at the VLA using the 6 cm and 20 cm
bands. The 20 cm observations were performed in the B array
configuration, with 25 MHz bandwidth, while the 6 cm observations were
performed in the C array configuration, with 50 MHz bandwidth. The
sources were observed at two frequencies within each band in order to
obtain adequate coverage in $\lambda^2$ and determine the RM
unambiguously. This setup allows one to have the same UV-coverage in
both bands and offers the best compromise between resolution and
sensitivity to the sources' extended structure. The resulting angular
resolution of $\sim$4$''$ corresponds to $\sim$ 2 kpc at the cluster's
redshift. Having a high resolution is crucial to determine small-scale
RM fluctuations. At the same time, we also need good sensitivity to the
extended emission, in order to image RM variations on the largest
scales. The largest angular scale (LAS) visible in the 20-cm band with the B
array is 120$''$. From NVSS the sources 5C4.20 and 5C4.43 have a
larger angular extent, hence we also observed them with C array
configuration. Details of the observations are given in Table
\ref{tab:radioobs}.  Since observations were taken in the VLA-EVLA
transition period, baseline calibration was performed, using the
source 1310+323 as calibrator. The source 3C286 was used as both
primary flux density calibrator\footnote{we refer to the flux density
  scale by \citep{Baars77}} and as absolute reference for the electric
vector polarisation angle. The source 1310+323 was observed as both a
phase and parallactic angle calibrator. \\ We performed standard
calibration and imaging using the NRAO Astronomical Imaging Processing
Systems (AIPS). Cycles of phase self-calibration were performed to
refine antenna phase solutions on target sources, followed by a final
amplitude and gain self-calibration cycle in order to remove minor
residual gain variations.  Total intensity, I, and Stokes parameter Q
and U images have been obtained for each frequency separately. The
final images were then convolved with a Gaussian beam having FWHM $=$
5$''\times$5$''$ ($\sim$ 2.3$\times$2.3 kpc). Polarization intensity
$P=\sqrt{U^2+Q^2}$, polarization angle $\Psi=\frac{1}{2}\atan(U,Q)$
and fractional polarization $FPOL=\frac{P}{I}$ images were obtained
from the I, Q and U images. Polarization intensity images have been
corrected for a positive bias. The calibration errors on the measured
fluxes are estimated to be $\sim$ 5\%. \\

\begin{table*} 
\caption{VLA observations of radio galaxies in the Coma cluster field.}        %
\label{tab:radioobs}      
\centering          
\begin{tabular}{|c c c c c c c c|}     
\hline\hline       
Source       & RA           &  DEC      &  $\nu$    & Bandwidth & Config. & Date  & Net Time on Source   \\
             & (J2000)      & (J2000)   &   (GHz)   & (MHz)     &         &       & (Hours)  \\\hline                    
5C4.20 &  12 54 18.8  & +27 04 13       & 4.535 - 4.935   & 50  &   C     & Jul 09&  3.0   \\ 
       &              &                 & 1.485 - 1.665   & 25  &   B     & May 09&  1.6  \\
       &              &                 & 1.485 - 1.665   & 25  &   C     & Jul 09&  1.6  \\
\hline
5C4.24 &  12 54 58.9  & +27 14 51       & 4.535 - 4.935   & 50  &   C     & Jul 09&  3.5   \\ 
       &              &                 & 1.485 - 1.665   & 25  &   B     & May 09&  2.0  \\
\hline
5C4.29 &  12 55 21.0  & +27 14 44       & 4.535 - 4.935   & 50  &   C     & Jul 09&  3.0  \\ 
       &              &                 & 1.485 - 1.665   & 25  &   B     & May 09&  2.0  \\    
\hline
5C4.31 &  12 55 43.3  & +27 16 32       & 4.535 - 4.935   & 50  &   C     & Jul 09&   3.2  \\ 
       &              &                 & 1.485 - 1.665   & 25  &   B     & Apr - May 09&  3.2  \\   
\hline
5C4.43 &  12 56 43.6  & +27 10 41       & 4.535 - 4.935   & 50  &   C     & Jul 09      &  3.0   \\ 
       &              &                 & 1.485 - 1.665   & 25  &   B     & Apr - May 09&  2.8  \\   
       &              &                 & 1.485 - 1.665   & 25  &   C     & Jul 09       & 1.6   \\
\hline
5C4.51 &  12 57 24.3  & +27 29 51       & 4.535 - 4.935   & 50  &   C     & Jul 09& 3.0    \\ 
       &              &                 & 1.485 - 1.665   & 25  &   B     & May 09& 3.2   \\   

\hline                  
\multicolumn{8}{l}{\scriptsize Col. 1: Source name; Col. 2, Col. 3: Pointing position (RA, DEC);
Col. 4: Observing frequency;}\\
\multicolumn{8}{l}{\scriptsize Col 5: Observing bandwidth; Col. 6: VLA configuration; 
Col. 7: Dates of observation;}\\
\multicolumn{8}{l}{\scriptsize Col. 8: Time on source (flags taken into account).}\\ 
\end{tabular}
\end{table*}
\begin{table*}
\caption{ Total and polarization intensity radio images. Images are restored with a beam of 5$''\times$5$''$ }
\label{tab:radiomaps}
\centering

\begin{tabular} {|c c c c c c c c|} 
\hline\hline
Source name & $\nu $  &   $\sigma$(I) & $\sigma$(Q) & $\sigma$(U) & Peak brightness  &Flux density & Pol. flux \\
             & (GHz)          & (mJy/beam)   & (mJy/beam)  & (mJy/beam)  & (mJy/beam)         & (mJy)       & (mJy)     \\ 
\hline
5C4.20  &  1.485    & 0.050 & 0.024 & 0.024  &9.0206E-03   &  8.1052E-02   & 1.6721E-02   \\
        &  1.665    & 0.059 & 0.026 & 0.026  &9.2810E-03   &  8.0599E-02   & 1.7609E-02 \\
        &  4.535    & 0.020 & 0.017 & 0.017  &5.5346E-03   &  4.0829E-02   & 8.8433E-03  \\
        &  4.935    & 0.019 & 0.018 & 0.018  &5.5642E-03   &  4.1919E-02   & 8.8699E-03   \\
\hline
5C4.16  &  1.485    & 0.050 & 0.024 & 0.024  & 2.1867E-02  & 4.2672E-02    & 2.7042E-03   \\
        &  1.665    & 0.059 & 0.026 & 0.026  & 2.1905E-02  & 3.9896E-02    & 2.7011E-03\\
        &  4.535    & 0.020 & 0.017 & 0.017  & 6.3046E-03  & 1.1521E-02    & 8.2612E-04  \\
        &  4.935    & 0.019 & 0.018 & 0.018  & 5.7240E-03  & 1.1033E-02    & 7.0368E-04  \\
\hline
5C4.24  &  1.485    & 0.048 & 0.025 & 0.024  &8.3123E-03  & 3.4E-02     &  1.4087E-03    \\
        &  1.665    & 0.049 & 0.029 & 0.028  &9.726E-04   & 2.8395E-02  &  1.8146E-03\\
        &  4.535    & 0.018 & 0.017 & 0.016  &4.4992E-03   &1.4653E-02  & 1.7374E-03   \\
        &  4.935    & 0.017 & 0.017 & 0.017  &4.4218E-03   &1.3855E-02  & 1.3685E-03  \\
\hline
5C4.29  &  1.485    & 0.049 & 0.023 & 0.023  &3.9605E-03   &1.6574E-02    & 7.4013E-04 \\
        &  1.665    & 0.049 & 0.028 & 0.028  &3.6253E-03   & 1.3414E-02   & 5.5392E-04\\
        &  4.535    & 0.018 & 0.017 & 0.017  & 2.7847E-03  & 6.2474E-03   & 1.9849E-04  \\
        &  4.935    & 0.018 & 0.018 & 0.017  & 2.7966E-03  & 5.1822E-03   & 1.9345E-04  \\
\hline
5C4.31  &  1.485    & 0.047 & 0.017 & 0.017  &7.4910E-03 &  1.9235E-02  & 1.9059E-03    \\
        &  1.665    & 0.046 & 0.023 & 0.022  &7.6109E-03 &  1.9606E-02  & 1.8874E-03\\
        &  4.535    & 0.017 & 0.017 & 0.017  &2.9999E-03 &  7.1626E-03  & 5.9231E-04  \\
        &  4.935    & 0.018 & 0.017 & 0.017  &2.9737E-03&   7.2758E-03 &  5.9357E-04 \\
\hline
5C4.43  &  1.485    & 0.045 & 0.019 & 0.021  & 8.5986E-03   & 5.0856E-02   & 7.2086E-03 \\ 
        &  1.665    & 0.041 & 0.023 & 0.023  &8.6323E-03   & 4.5734E-02   & 7.2029E-03\\   
        &  4.535    & 0.015 & 0.013 & 0.013  & 6.1534E-03   & 2.5576E-02  & 3.0802E-03  \\ 
        &  4.935    & 0.015 & 0.013 & 0.014  & 6.2770E-03   & 2.5461E-02   & 2.8089E-03  \\  
\hline
5C4.51  &  1.485    & 0.051 & 0.022 & 0.021  & 8.3728E-03  & 6.6565E-02    &4.1651E-03     \\
        &  1.665    & 0.058 & 0.024 & 0.024  & 8.1401E-03  & 6.4310E-02    &4.0893E-03 \\
        &  4.535    & 0.018 & 0.017 & 0.017  & 3.8679E-03  & 2.7415E-02   & 2.1236E-03  \\
        &  4.935    & 0.018 & 0.017 & 0.018  & 3.6772E-03  & 2.5692E-02   & 2.0066E-03  \\
\hline

\hline \multicolumn{8}{l}{\scriptsize Col. 1: Source name;
  Col. 2: Observation frequency; Col. 3, 4, 5:
  RMS noise of the I, Q, U images; }\\ \multicolumn{8}{l}{\scriptsize
  Col. 7: Peak brightness; Col. 8: Flux density; Col. 9: Polarized
  flux density.}
\end{tabular}
\end{table*}

\subsection{Radio properties of the observed sources}
\label{sec:radioSou}
In this section the radio properties of the observed sources are
briefly presented. Further details are given in Table
\ref{tab:radiomaps}.\\ Redshift information is available for three out
of the seven radio sources.  Although the redshift is not known
for the other four radio sources, they have not been associated with
any cluster galaxy down to very faint optical magnitudes: M$_r \geq$
-15 (see Miller et al. 2009).  This indicates that they are background
radio sources, seen in projection through the radio relic. In the
following, the radio emission arising from the selected sample of
sources is described together with their main polarisation properties.\\
\smallskip\\ {\bf 5C4.20 - NGC 4789}\\ The radio emission of NGC 4789
is associated with an elliptical galaxy with an apparent optical
diameter of $\sim$ 1$'$.7 located at redshift z$\sim$0.028 (De Vacoulers et al. 1976). It lies at $\sim
1.5^{\circ}$ from the cluster centre, South-West of the Coma
relic. The radio source is characterised by a Narrow Angle Tail (NAT)
structure.  In our high-resolution images the source shows two
symmetric and collimated jets that propagate linearly from the centre
for $\sim 35''$ in the SE- NW direction (see
Fig. \ref{fig:20_RM}). Then, the jets start bending toward North-East
up to a linear distance of $\sim 130''$ from the galaxy.  The
brightness decreases from the centre of the jets towards the lobes
that appear more extended in the 20-cm band images. On average the
source is polarised at the 20\% level at 1.485 GHz and at the 24\%
level at 4.935 GHz.  Lower resolution images by
\citet{1989A&A...213...49V} show that the total extent of the source
is $\sim 6'$, from the core to the outermost low-brightness
features. \citet{1989A&A...213...49V} also note that no extended lobes
are present at the edges of the jets, and the morphology of the low
brightness regions keeps following the jets' direction without
transverse expansion.\\
{\bf  5C4.16}\\ Another source is located  3$'$.8 West of 5C4.20. It is {\bf
  5C4.16} (Fig. \ref{fig:16_RM}). 5C4.16 is moderately extended, with a LLS of
$\sim$20$''$, but enough to be resolved by more synthesised
beams. 5C4.16 is also polarised: 7\% on average at both 1.485 GHz and
5.985 GHz.  Hence, although this source was not the target of the
observation, it adds a piece of information to our analysis.\\ {\bf
  5C4.24}\\ The radio emission from the source is likely associated
with the lobe of a radio galaxy centred on RA$=$12h55m08.3s and
DEC$=$+27d15m34s (see Fig. \ref{fig:24_RM}). The redshift of the
source is $z=$0.257325, according to
\citet{2002ApJS..143....1M}. Hence, 5C4.24 lies in the background of
the Coma cluster and  is projected onto the radio relic. In our
images, we detect the nucleus, the lobe and a weak counter-lobe (which is not shown in Fig. \ref{fig:24_RM}). The
projected position of 5C4.24 is at $\sim 1.2^{\circ}$ from the Coma
cluster centre. The emission from the counter-lobe is too weak to be
detected in polarisation. However, the lobe shows a mean fractional
polarisation of 8\% at 1.485 GHz, which increases to 13\% at 4.935
GHz. \\ {\bf 5C4.29}\\ 5C4.29 is located at $\sim 1.2^{\circ}$ from
the centre of the Coma cluster.  Three distinct components are visible
in our images (Fig. \ref{fig:29_RM}). A nucleus and two bright lobes
are detected from 1.485 GHz to 4.935 GHz. The western lobe is
connected to the nucleus through a region of lower brightness, visible
only in the 20-cm band images. No redshift is available for this
radio-source. Its total extension is $\sim 68''$. We note that if the
source was located at the Coma redshift, this would translate into a
linear size of 31 kpc, that would be exceptionally small for a
radio-galaxy.  The western lobe is polarised on average at the 7\%
level at both 1.485 GHz and 4.935 GHz, while the eastern lobe is
polarised at 5\% and 8\% at 1.485 GHz and 4.935 GHz respectively.\\
{\bf 5C4.31}\\ Our images suggest that the source 5C4.31 is associated
with the brightest lobe of a radio galaxy (see
Fig. \ref{fig:31_RM}). A weaker counter-lobe is also visible in our
images, while the nucleus is only detected in the 6-cm band images. It is located in projection onto
the radio relic, close to its eastern edge. The projected distance
from the cluster centre is $\sim 1.1^{\circ}$. The lobe exhibits a
mean fractional polarisation of $\sim$13\% and 10\% at 1.485 and 4.935
GHz, respectively. \\ {\bf 5C4.43 - NGC 4827}\\ The radio emission
from 5C4.43 is associated with the elliptical galaxy NGC4827 at z$=$ 0.025 \citep{2000MNRAS.313..469S}. It is located in-between the Coma cluster and
the radio relic, at $\sim 1^{\circ}$ from the cluster centre. The
radio source was studied in detail by \citet{1989A&A...213...49V}: it has a Z-shaped morphology with some evidence of jets
joining the core of the radio source to the diffuse emission of the
lobes. The angular extent of the source at 1.4 GHz is $\sim 4'30''$
\citep{1989A&A...213...49V}.  Our observations show the core and two
bright jets departing in the N and S direction. The brightness of the jets
decreases as they propagate out from the nucleus, and fade into two
diffuse lobes (Fig. \ref{fig:43_RM}). The lobes appear more extended
in our 20-cm band images. This is due to the steep spectrum that
characterises the lobes or the radio-galaxies. Both, the
jets and the diffuse lobes are polarised. We detect a mean
polarisation of $\sim$10\% at 1.485 GHz and 4.935 GHz.\\ {\bf
  5C4.51 - NGC 4839 }\\ The radio source 5C4.51 is identified 
  with NGC 4839, the brightest galaxy of the group NGC 4839, which is currently merging with the Coma
cluster. The redshift of the source is $\sim$ 0.25 \citep{2000MNRAS.313..469S}.
5C4.51 is located at $\sim 0.6^{\circ}$ from the Coma cluster
centre.  The source largest extension is $\sim 1'$, and it is composed
of two extended and connected radio lobes, while the nucleus is not
resolved. The source is polarised on average at the 7\% and 9\% level
at 1.485 GHz and
4.935 GHz, respectively. \\

\section{Rotation Measures}
\label{sec:RMobs}
\begin{figure*}
\centering
\includegraphics[width=0.8\textwidth]{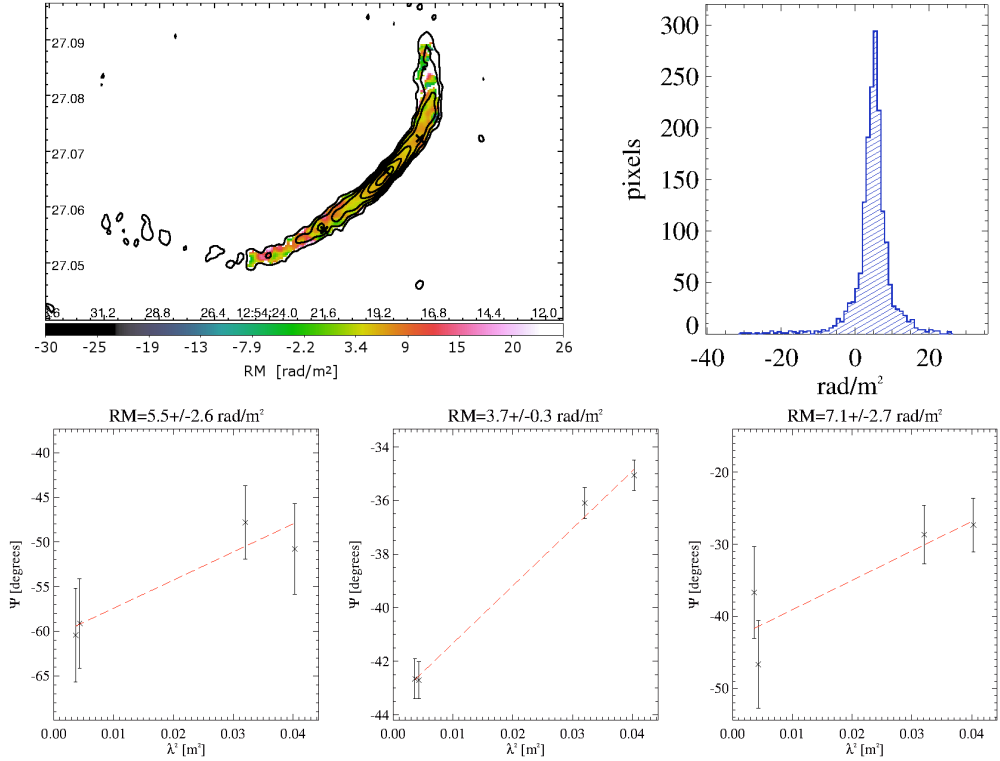}
\caption{{\bf 5C4.20:} {\it Top left:} The RM fit is shown in colour along
with total intensity radio
  contours 1.4 GHz. The bottom contour correspond to the 3$\sigma$
  noise level and contours are then spaced by a factor of 2. {\it Top
    right:} distribution histogram of the RM values. {\it Bottom:} fits of
  polarisation angle versus $\lambda^2$ in three representative
  pixels marked with crosses in the top-left image. }
\label{fig:20_RM}
\end{figure*}

\begin{figure*}
\centering
\includegraphics[width=0.8\textwidth]{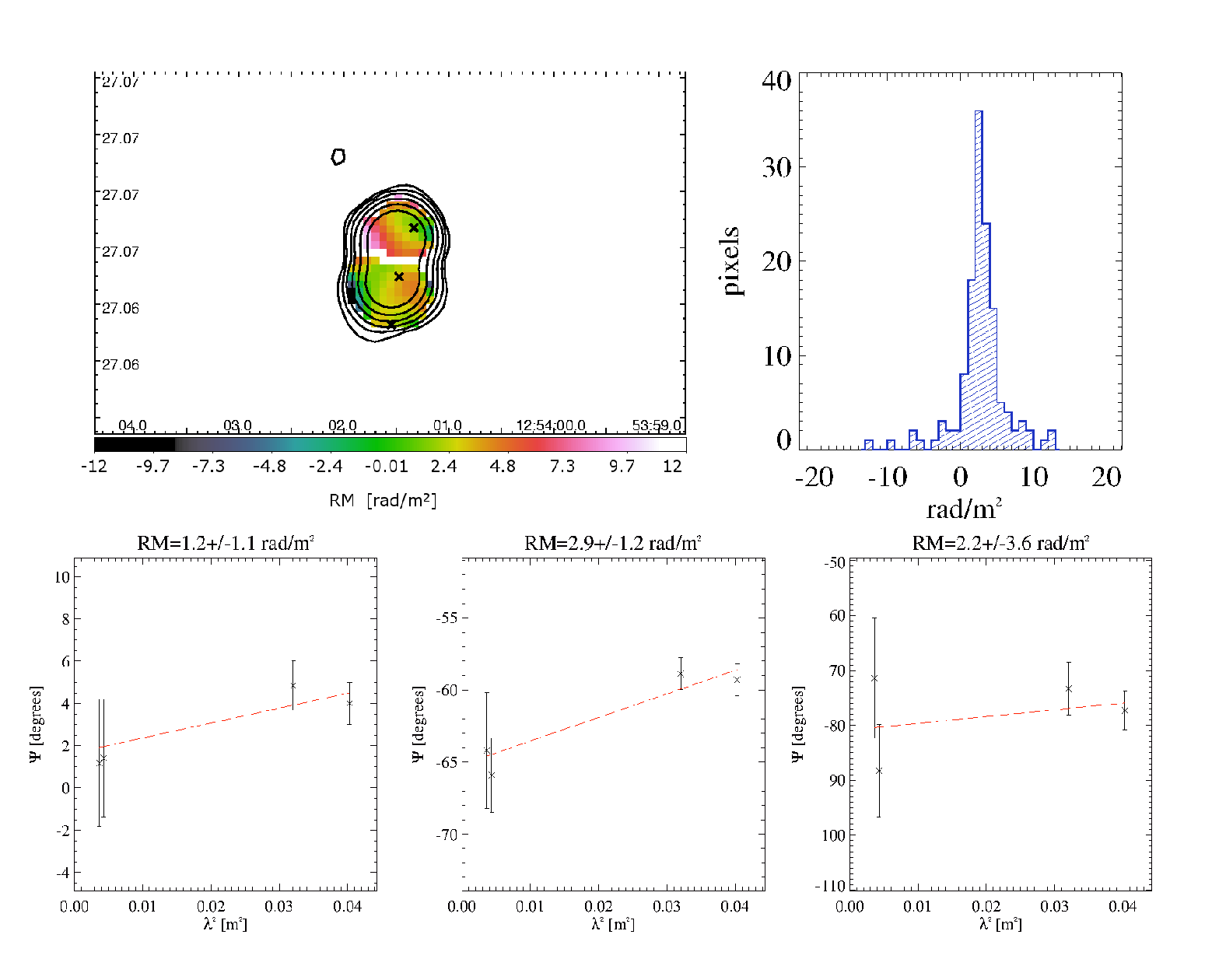}
\caption{{\bf 5C4.16:} {\it Top left:} The RM image is shown in color along with total
  intensity radio contours at 1.4 GHz. Contours start at 3$\sigma$ and 
increase by factors of 2. {\it Top right:} distribution histogram of the RM
  values. {\it Bottom:} fits of polarisation angle versus $\lambda^2$ in
  three representative pixels marked with crosses in the top-left image. }
\label{fig:16_RM}
\end{figure*}

\begin{figure*}
\centering
\includegraphics[width=0.8\textwidth]{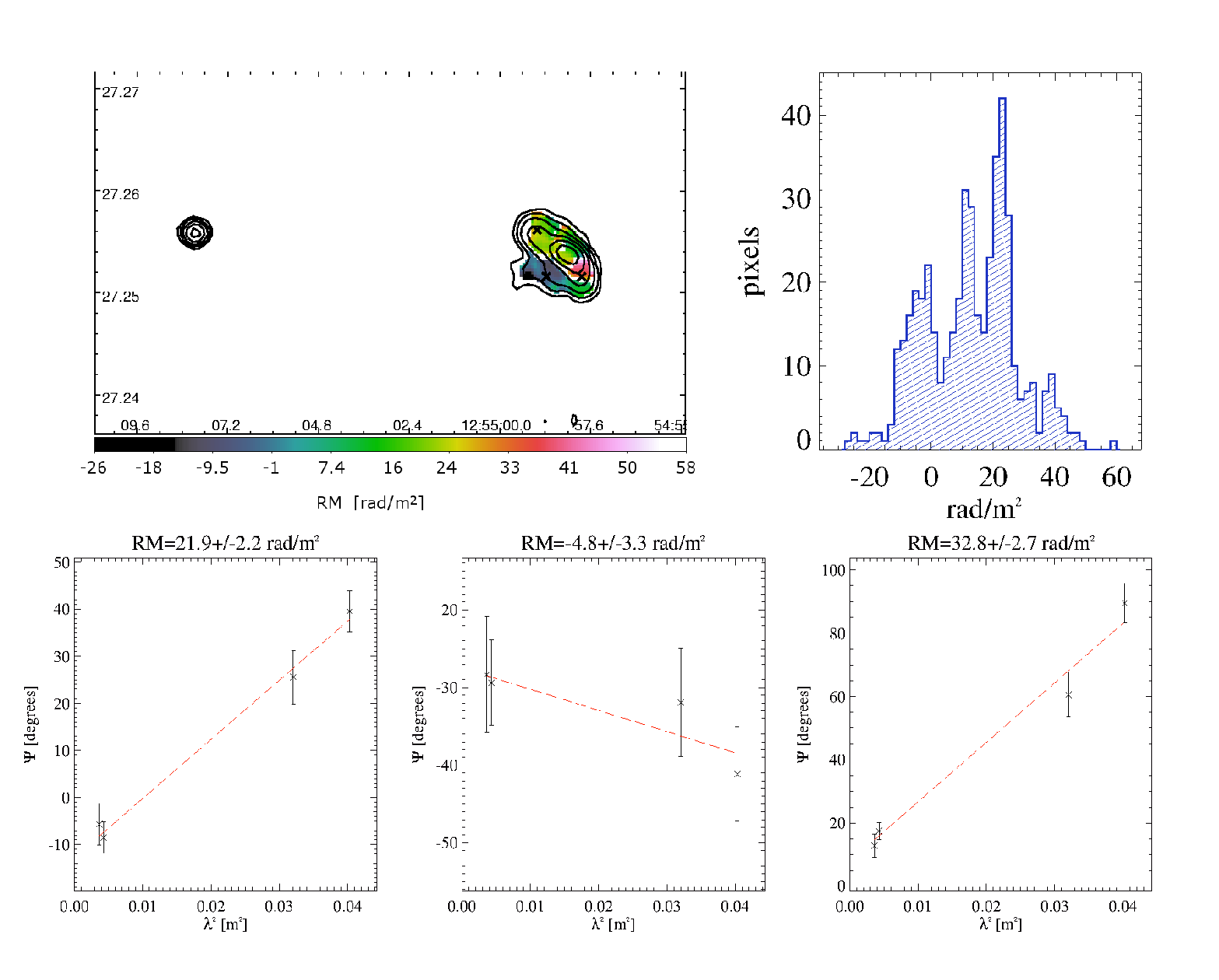}
\caption{{\bf 5C4.24:} {\it Top left:} The RM image is shown in colour along
  with total intensity radio contours at 1.4 GHz. Contours start at 3$\sigma$ and 
increase by factors of 2. {\it Top right:} distribution histogram of the RM values.
  {\it Bottom:} fits of polarisation angle versus $\lambda^2$ in three
  representative pixels marked with crosses in the top-left image. }
\label{fig:24_RM}
\end{figure*}

\begin{figure*}
\centering
\includegraphics[width=0.8\textwidth]{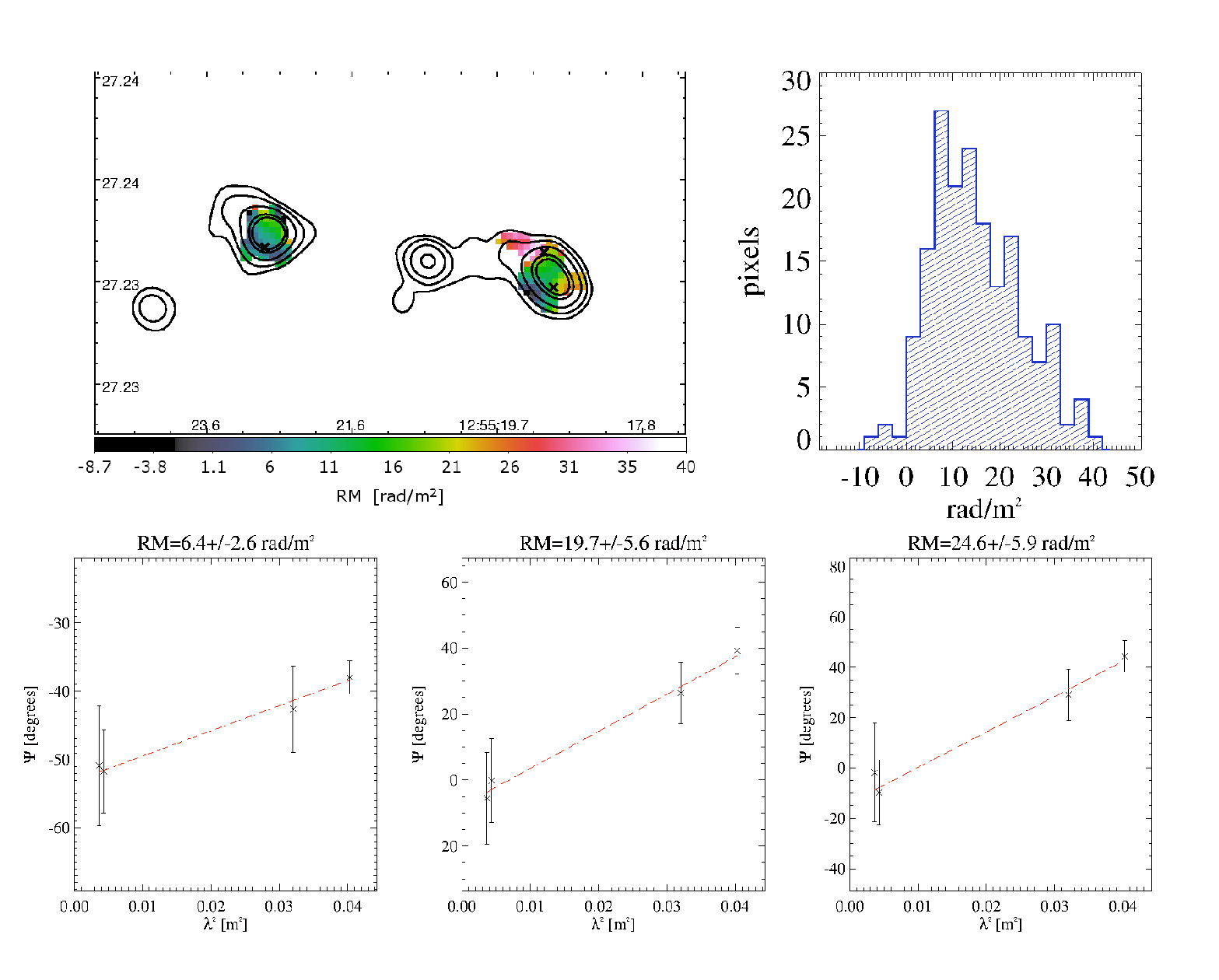}
\caption{{\bf 5C4.29:} {\it Top left:} The RM image is shown in colours along
  with total intensity radio contours at 1.4 GHz. Contours start at 3$\sigma$ and 
increase by factors of 2.  {\it Top
    right:} distribution histogram of the RM values. {\it Bottom:} fits of
  polarisation angle versus $\lambda^2$ in three representative
  pixels marked with crosses in the top-left image. }
\label{fig:29_RM}
\end{figure*}

\begin{figure*}
\centering
\includegraphics[width=0.8\textwidth]{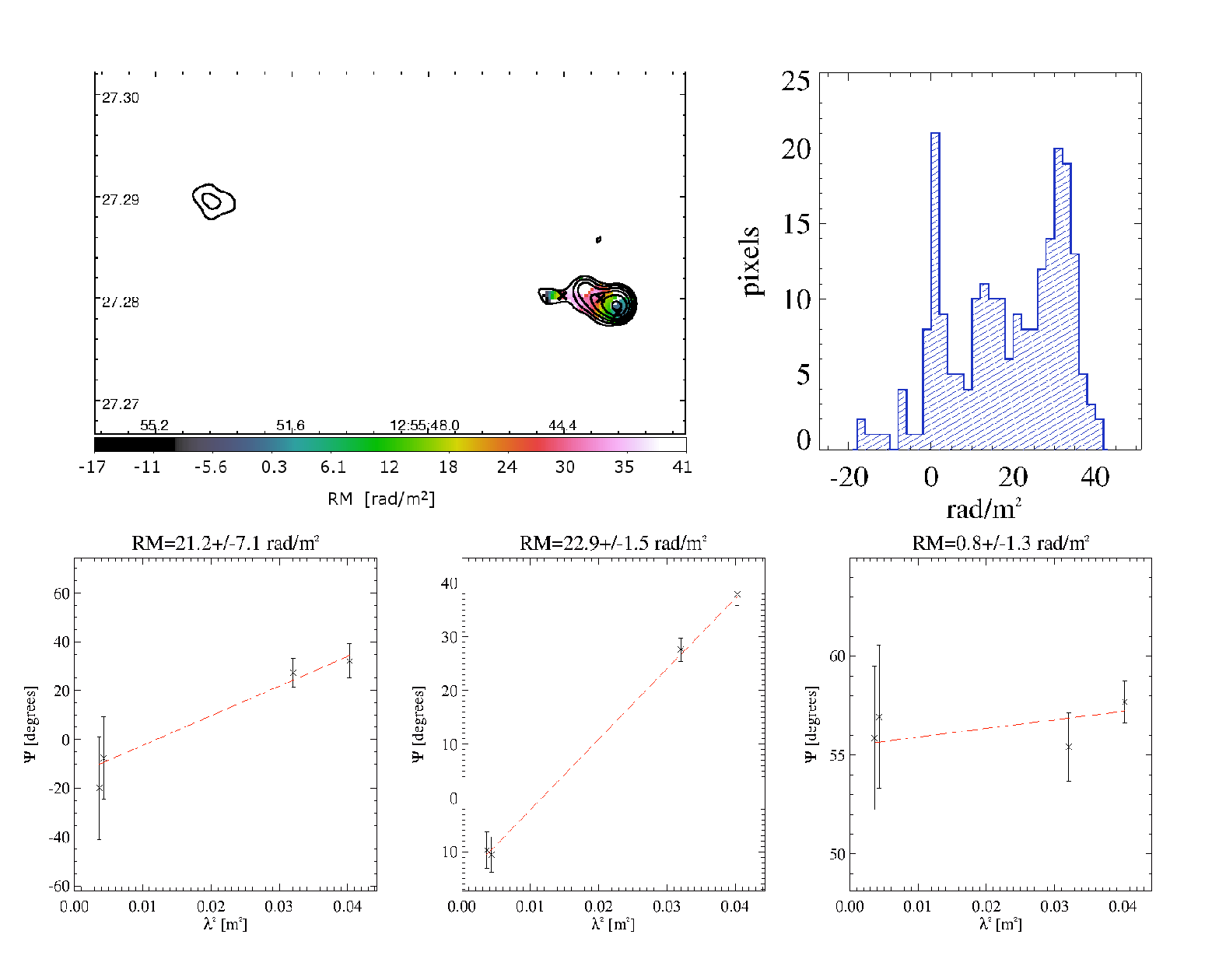}
\caption{{\bf 5C4.31:} {\it Top left:} The RM image is shown in colour
  along with total intensity radio contours at 1.4 GHz. 
Contours start at 3$\sigma$ and 
increase by factors of 2. {\it Top right:} distribution histogram of the RM values.
  {\it Bottom:} fits of polarisation angle versus $\lambda^2$ in three
  representative pixels marked with crosses in the top-left image. }
\label{fig:31_RM}
\end{figure*}

\begin{figure*}
\centering
\includegraphics[width=0.8\textwidth]{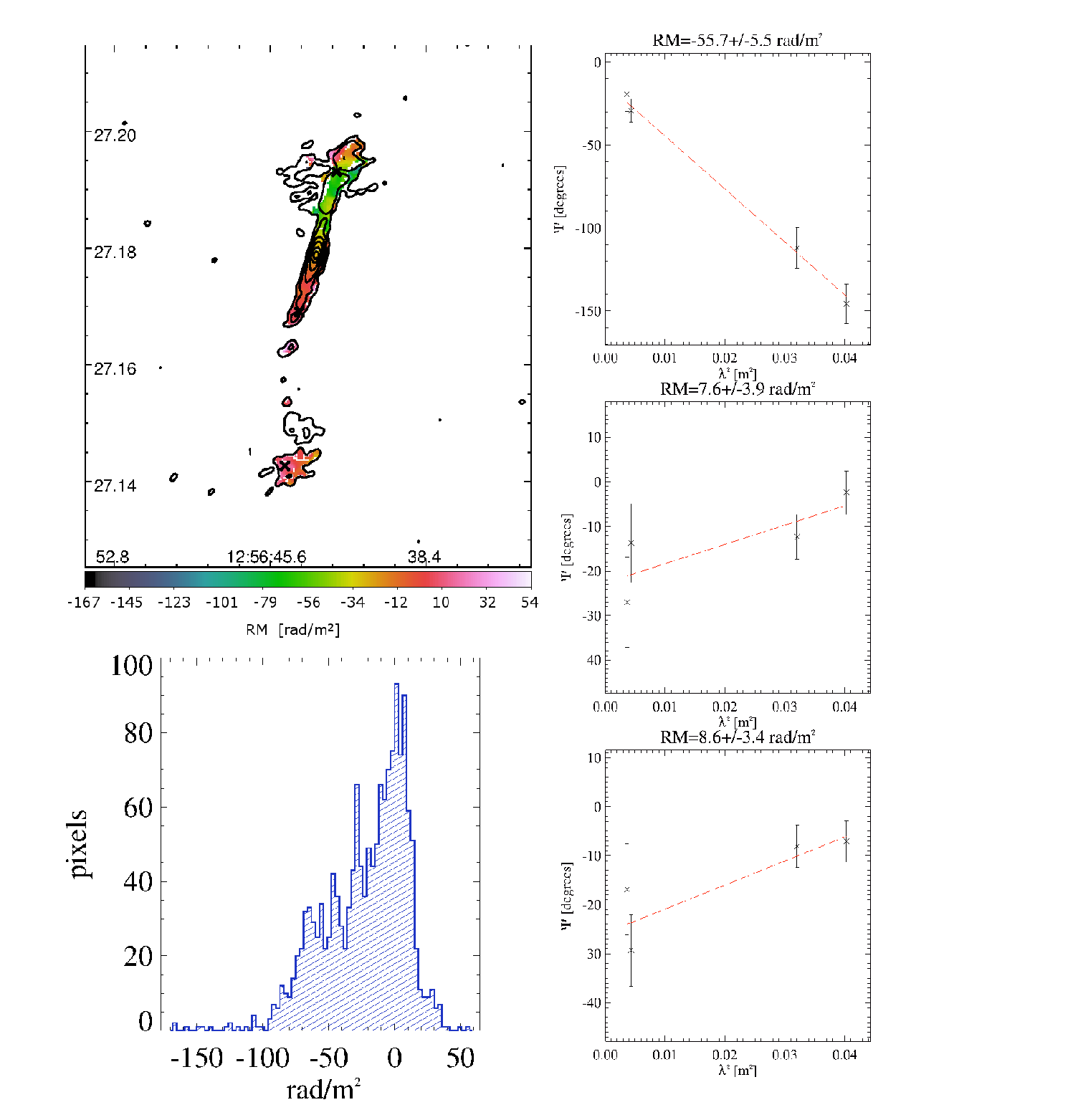}
\caption{{\bf 5C4.43:} {\it Left top:} The RM image is shown in colour 
  along with total intensity radio contours at 1.4 GHz. 
Contours start at 3$\sigma$ and 
increase by factors of 2.
{\it Left bottom:} distribution histogram of the RM values.
  {\it Right:} fits of polarisation angle versus $\lambda^2$ in three
  representative pixels marked with crosses in the left-top image. }
\label{fig:43_RM}
\end{figure*}

\begin{figure*}
\centering
\includegraphics[width=0.8\textwidth]{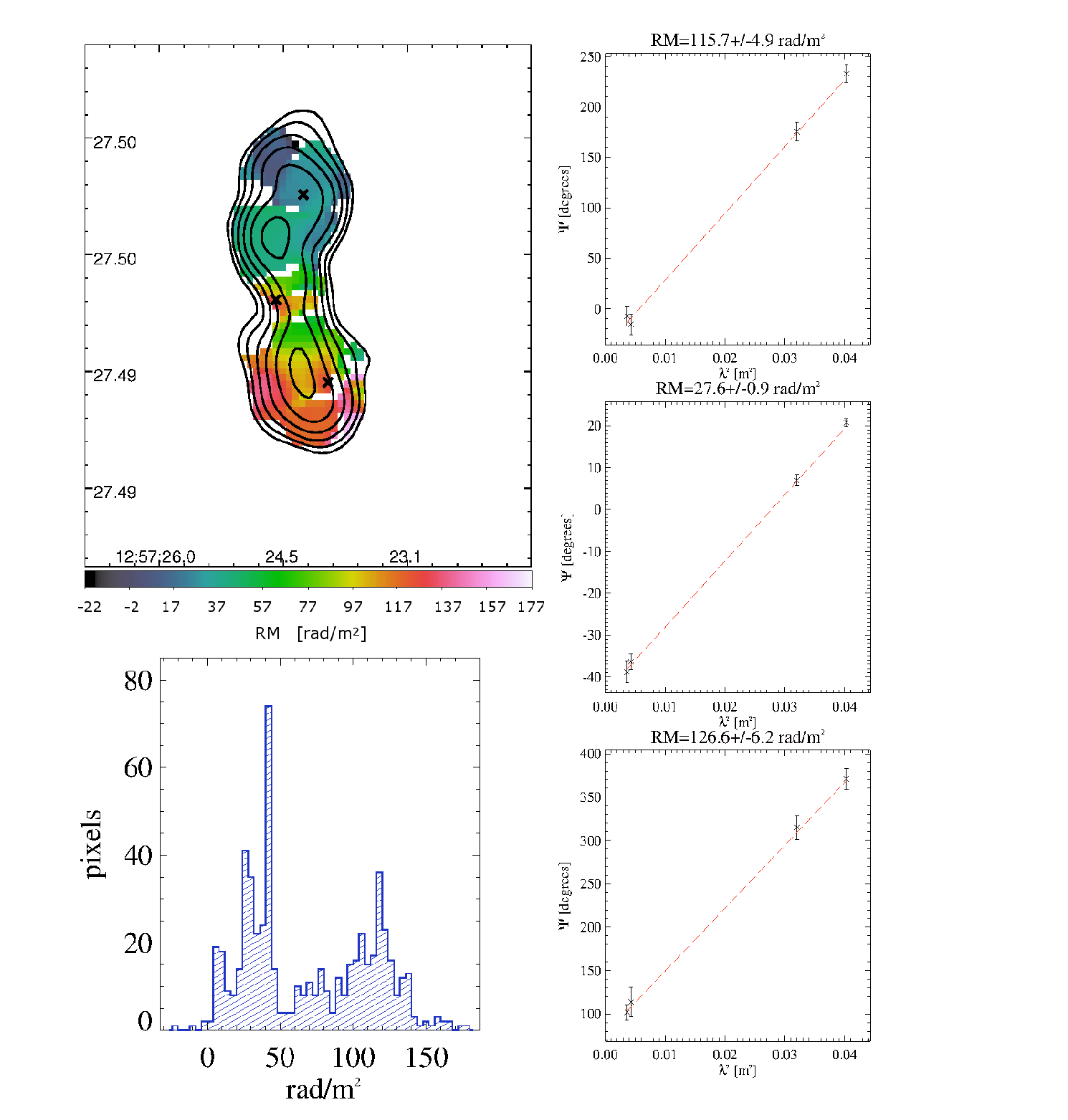}
\caption{{\bf 5C4.51:} {\it Left top:} The RM image is shown in colour along
with total intensity radio contours at 1.4 GHz. 
Contours start at 3$\sigma$ and 
increase by factors of 2. {\it Left bottom:} distribution histogram of the RM values. {\it Right:}
  fits of polarisation angle versus $\lambda^2$ in three representative
  pixels marked with crosses in the left-top image. }
\label{fig:51_RM}
\end{figure*}

We derived the fit to the RMs from the polarisation angle images using
the PACERMAN algorithm (Polarization Angle CorrEcting Rotation Measure
ANalysis) developed by \citet{PACERMAN}. The algorithm uses a few
selected reference pixels to solve for the n$\pi$-ambiguity in a
nearby area of the map. Once the n$\pi$-ambiguity is solved for, the
RMs can be computed also in low signal-to-noise regions.  As reference
pixels we considered those with a polarisation angle uncertainty of less
than 10 degrees, corresponding to 2$\sigma$ level in both U and Q
polarisation maps simultaneously. The n$\pi$ value found for the
reference pixel is then transferred to the nearby pixels, if their
polarisation angle gradient is below a certain threshold at all
frequencies. We fixed this threshold to 15 to 25
degrees, depending on the image features.  The resulting RM images are
shown in Fig. \ref{fig:20_RM}, \ref{fig:16_RM}, \ref{fig:24_RM},
\ref{fig:29_RM}, \ref{fig:31_RM}, \ref{fig:43_RM}, and \ref{fig:51_RM}
overlaid onto the total intensity contours at 1.485 GHz. In the same
figures, we also display the RM distribution histograms and the RM fits
for some representative pixels, marked by crosses in the respective
images.\\ The polarisation angle $\Psi$ follows the $\lambda^2$-law,
expected in the case of an external Faraday screen. We will assume in
the following that the Faraday Rotation is occurring entirely in the
ICM and it is hence representative of the magnetic field weighted by
the gas density of the cluster.  We will discuss this assumption in
Sec. \ref{sec:RMlocal}.\\ From the RM images, we computed the RM mean
($\langle \rm RM\rangle$) and its dispersion ($\sigma_{\rm
  {RM,obs}}$) (see Table \ref{tab:rm1}).

\begin{table}
\caption{Rotation Measure values of the observed sources}          
\label{tab:rm1}      
\centering          
\begin{tabular}{|c c c c c c|}    
\hline
\hline
Source    & Proj. distance & n. of beams &$\langle RM\rangle$    &$\sigma_{\rm{RM}}$& Err$_{fit}$ \\
          &   kpc    &     & rad/m$^2$   & rad/m$^2$       &  rad/m$^2$   \\    
\hline            
5C4.20    &   2451   &  62  &4.8 $\pm$0.7  &  5.2$\pm$0.6    &  2.9    \\
5C4.16    &   2543   &   5  &2.8 $\pm$2    &  3.3$\pm$1.5    &  1.5    \\ 
5C4.24    &   2075   &  16  &13  $\pm$4    &  15 $\pm$3      &  4.3    \\ 
5C4.29    &   1982   &   7  &15  $\pm$4    &  10 $\pm$4      &  5.8    \\
5C4.31    &   1824   &   8  & 18 $\pm$5    &  14 $\pm$4      &  3.2      \\
5C4.43    &   1667   &  55  & -22 $\pm$4   &  31 $\pm$3      &   5.5     \\      
5C4.51    &   1113   &  21  & 69 $\pm$ 9   & 43 $\pm$ 7      &   3.9     \\  
\hline
\multicolumn{6}{l}{\scriptsize Col. 1: Source name Col. 2: Source  projected distance from the X-ray cluster centre;}\\
 \multicolumn{6}{l}{\scriptsize Col. 3:  number of beams over which RMs are computed;}\\
\multicolumn{6}{l}{\scriptsize Col. 4: Mean value of the RM distribution;}\\
\multicolumn{6}{l}{\scriptsize Col. 5: Dispersion of the RM distribution; }\\
\multicolumn{6}{l}{\scriptsize  Col. 6: Median of the RM fit error.}
\end{tabular}
\end{table}

\subsection{Errors on the RM}
The errors that affect the values of $\langle \rm RM\rangle$ and
$\sigma_{\rm{RM,obs}}$ can be separated into fit errors and statistical
errors.  The former has the effect of increasing the {\it real} value
of $\sigma_{\rm{RM}}$, while the latest is due to the finite sampling of
the RM distribution. In our case, it is determined by the number of
beams, $n_b$, over which the RM is computed.  In order to derive the
{\it real} standard deviation of the observed RM distribution, we have
computed the $\sigma_{\rm{RM}}$ as
$\sqrt{\sigma^2_{\rm{RM,obs}}-\rm{Err}_{\rm{fit}}^2}$. $\rm{Err_{fit}}$ is the median of
the error distribution, and $\sigma^2_{\rm{RM,obs}}$ is the observed value
of the RM dispersion.  Errors have been estimated with Monte Carlo
simulations, following the same approach used in
\citet{Bonafede10}.  We have extracted $n_b$ values from a
random Gaussian distribution having $\sigma=\sigma_{\rm{RM,obs}}$ and mean
$=\langle \rm RM \rangle$. 
In order to mimic
the effect of the noise in the observed RM images, we have added to the extracted values
Gaussian noise having $\sigma_{\rm{noise}}=Err_{\rm{fit}}$. We have computed
the mean and the dispersion ($\sigma_{\rm{sim}}$) of these simulated
quantities and then subtracted the noise from the dispersion obtaining
$\sigma_{\rm{sim,dec}}=\sqrt{\sigma_{\rm{sim}}^2-\sigma_{\rm{noise}}^2}$. 
Thus, we have obtained a distribution of $\sigma_{\rm{sim,dec}}$ and means.
The standard deviation of the $\sigma_{\rm{sim,dec}}$ distribution is
then assumed to be the uncertainty on $\sigma_{\rm{RM,dec}}$, while the standard deviation of the
mean distribution is assumed to be  the fit error on $\langle RM\rangle$. We checked
that the mean of both distributions recover the corresponding observed
values.  In Table \ref{tab:rm1} we list the RM mean, its dispersion
($\sigma_{\rm{RM}}$), with the respective errors, the median of the fit
error ($\rm{Err}_{\rm{fit}}$), and the number of beams over which the RM
statistic is computed ($n_b$).

\subsection{Galactic contribution} 
The contribution to the Faraday RM from our Galaxy may introduce an
offset in the Faraday rotation that must be removed.  This
contribution depends on the Galactic positions of the observed
sources. The Galactic coordinates of the Coma cluster are
$l=58^{\circ}$ and $b=88^{\circ}$. The cluster is close to the
Galactic north pole, so that Galactic contribution to the observed RM
is likely negligible.  We have estimated the Galactic contribution
as in \citet{Bonafede10}. Using the catalogue by
\citet{Simard81}, the average contribution for extragalactic sources
located in projection close to the Coma cluster is $\sim$
-0.15 rad/m$^2$ in a region of 25$\times$25 degrees$^2$ centered on
the cluster (The RM from each source has been weighted by the inverse
of the distance from the centre of the Coma cluster).  This small
contribution is  negligible and  will not be considered in the following analysis. 

\subsection{RM local contribution}
\label{sec:RMlocal}
\begin{figure*}
\centering
\includegraphics[width=\textwidth]{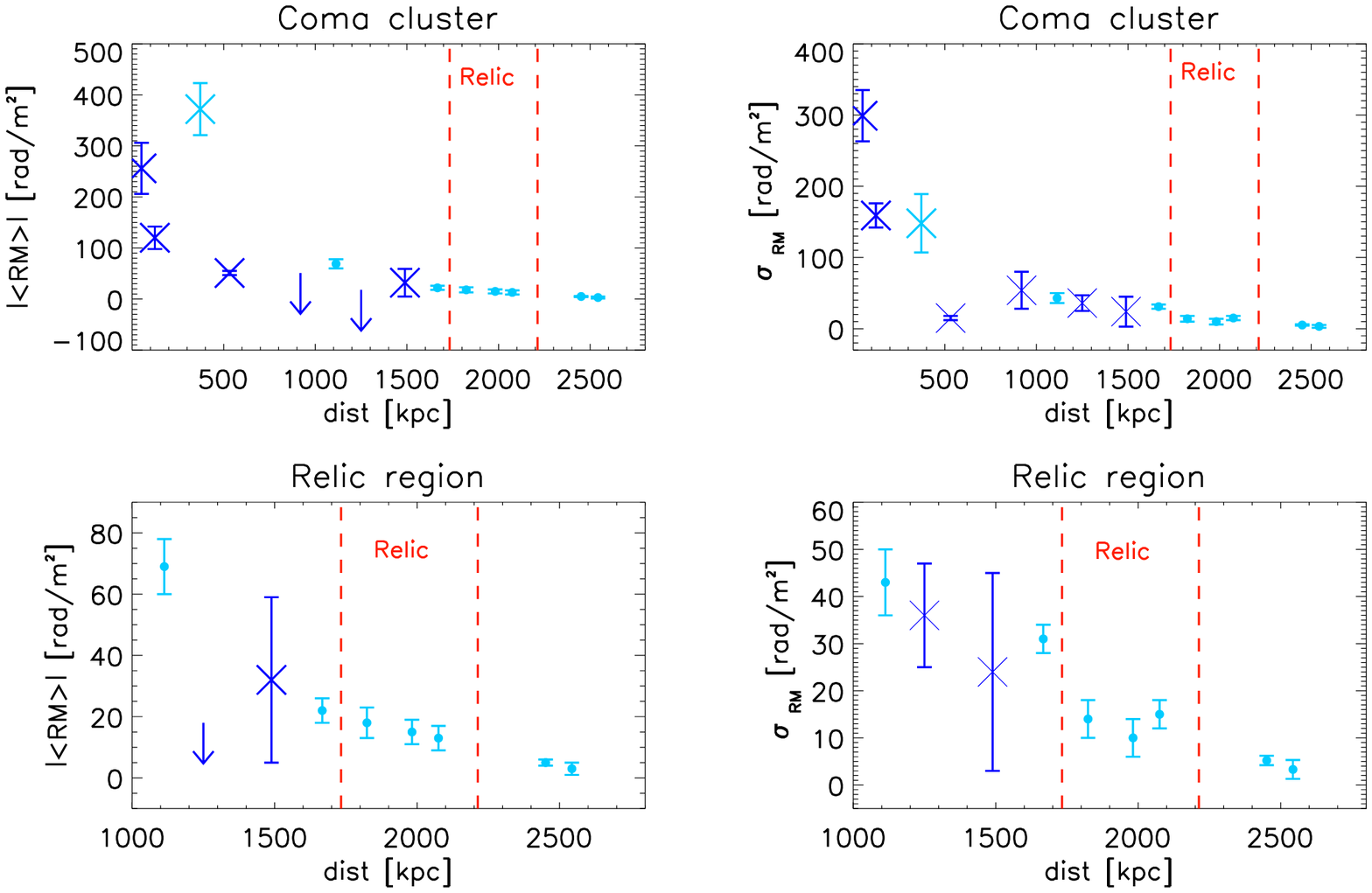}
\caption{ $\langle |\rm{RM}| \rangle$ and $\sigma_{\rm{RM}}$ trend versus the
  cluster projected distance.  Crosses are the points from
  \citet{Bonafede10}, dots are the new points from this paper.  Blue
  symbols mark the sources in the Coma cluster, cyan symbols mark
  those SW from the cluster centre, in the direction of the Coma
  relic. Red vertical lines indicate the approximative location of the
  Coma relic. In the top panels all the sources and displayed, in the bottom
  panels a zoom in the SW region is shown.}
\label{fig:rmobs}
\end{figure*}

 Here we discuss the possibility that the RMs observed in radio
 galaxies are not associated with the foreground ICM but may arise
 locally to the radio source (Bicknell et al. 1990, Rudnick \&
 Blundell 2003, see however Ensslin et al. 2003), either in a thin
 layer of dense warm gas mixed along the edge of the radio emitting
 plasma, or in its immediate surroundings.  Recently,
 \citet{Guidetti12} and \citet{2013arXiv1301.1400O} have found a local
 contribution to the RM in the lobes of B2 0755+37 and CentaurusA
 smaller than 20 rad/m$^2$, much smaller than the values measured in
 the Coma radiogalaxies analysed in \citet{Bonafede10}. This
   fact and the clear trend with the cluster projected distance
   indicate that the RM is due to the ICM rather than to local
   effectss. Although a local contribution cannot be rejected, it is
 unlikely to be dominating over the cluster ICM contribution. In
 addition, the RM map of B2 0755+37 shows stripes and gradients on the
 scales of the lobes that are pointing towards the presence of a local
 compression, while we do not detect any of such structures for the
 Coma cluster sources.  We speculate that if a local contribution was
 produced by compressed gas and field around the leading edges of the
 lobes, and if such contribution was the dominant one, then RM stripes
 or gradient on scales as large as the lobes should be detected in the
 RM map.  Additional statistical arguments against a local
 contribution are:
\begin{itemize}
\item{A trend of the RM versus the cluster impact parameter has been observed
    \citep[e.g.]{2004JKAS...37..337C,
      1999A&A...344..472F,2006A&A...460..425G,Bonafede11a};}\\
 \item{Statistical tests on the scatter plot of RM versus polarisation angle
 for several radio galaxies show that no evidence is found for a Faraday rotation local to radio lobes.\citep{2003ApJ...597..870E};}\\
\item{The relation between the maximum RM and the cooling flow rate in relaxed
 clusters \citep{2002MNRAS.334..769T} indicates that the RM of cluster sources is sensitive to the cluster environment  rather than to the lobe internal gas.}\\
\end{itemize}
 Among the sources analysed in this paper, only 3 belong to the 
 Coma cluster or NGC4839 group (namely, 5C4.20, 5C4.43, and 5C4.51),  while the other 4 are field sources in
 the background of the cluster.  
 Since there is no evidence for a local contribution in the RM images
 that we are presenting in this work (Fig. \ref{fig:20_RM} to \ref{fig:51_RM}), 
 and the trend of $\Psi$ versus $\lambda^2$ follows  the expectations
 for an external Faraday screen, we will assume that the RM we detect is originating 
 entirely in the ICM.\\
 The trend of $\langle |\rm{RM}| \rangle$ and $\sigma_{\rm RM}$ are
 shown in Fig. \ref{fig:rmobs} for the new sources analysed in this
 work as well as for the sources located in the Coma cluster
 \citep{Bonafede10}. The average trend indicates that the RM decreases
 from the centre towards the outskirts of the cluster, although an
 increment in both $\langle |\rm{RM}| \rangle$ and $\sigma_{\rm RM}$
 is detected in the SW quadrant. A net increase of the $\langle
 |\rm{RM}| \rangle$ is detected corresponding to the central sources
 of the group NGC 4839 interacting with the Coma cluster. In the
 following sections we will analyse if the detected trend is
 compatible with the magnetic field profile derived from the Coma
 cluster, or if an additional component of either the magnetic field, B, and/or  the gas density, $n$, is
 needed.

\section{Reproducing RM images}
\label{sec:code}
In this Section we describe the steps implemented in our method to
produce simulated trend of RM for galaxy clusters.  The numerical
procedures have been inspired by the work of Murgia et
al. (2004). The detailed implementation and developments, written in
IDL, will be subject of a forthcoming paper. We outline here
the main features of the code.\\ First, it produces a synthetic 
model for the gas density. Second, it creates a 3D distribution of a
magnetic field model from an analytical power spectrum within a fixed range of spatial scales, and scales it to
follow a given radial profile (or to scale with gas density).  Third,
it computes the RMs to be compared with real data taking into account
the real sampling of the observed maps. \\ In detail:\\ $\bullet$
{\bf Input gas density model.} We impose a model for the gas density,
like e.g. the  $\beta$-model \citep{betamodel} or a combination of $\beta$-models
 to reproduce non-isolated clusters. The input gas model could
also be taken directly from cosmological numerical simulations,
although we will not explore this in the present work.\\
$\bullet$ {\bf 3D cluster magnetic field.} 
In order to set up divergence-free, turbulent magnetic fields, we start
with the vector potential ${\bf A}$. In the Fourier domain, it is
assumed to follow a power-law with index $m$ between the minimum and
the maximum wavenumber $k_{\rm in}  \leq k \leq k_{\rm out}$:
\begin{equation}
P_{\rm A}(k)=| \rm{A_k}|^2 \propto k^{-m}.
\label{eq:pk}
\end{equation}
For all grid points in Fourier space, we extract random values of
amplitude and phase, corresponding to each component of $\bf{A_k}$.  In
order to obtain a random Gaussian distribution for ${\bf B}$, $\bf{A}$ is
randomly drawn from the Rayleigh distribution:
\begin{equation}
P(A, \phi)dA d\phi = \frac{A}{2\pi |A_k|^2} \exp \left(-\frac{A^2}{2 |A_k|^2} \right)  \cdot dA d\phi ,
\end{equation}
while $\phi$ is uniformly distributed in the range $[0,2\pi]$.
The combination of $A$ and $\phi$ allows us to compute the value of ${\bf {A_k}}$ for each point  in
the Fourier domain ($k_x$, $k_y$, $k_z$).
The magnetic field components in the Fourier space are then obtained
by computing the curl of  ${\bf { A_k}}$ thus generated:
\begin{equation}
{\bf {\tilde B}}(k) =i{\bf k}\times {\bf {A_k}}.
\end{equation}
The field components $B_i$ in real space are then derived using a 3D
Fast Fourier Transform.\\ The magnetic field
generated in this way is by definition divergence-free, with Gaussian
components $B_i$ having zero mean and a dispersion equal to
$\sigma_{B_i}=\langle B^2_i\rangle$. The magnetic field power spectrum is described
by a power law $|B_k|^2 \propto k^{-m+2}$.\\ In order to simulate a realistic profile
of cluster magnetic field, we need to scale $|{\bf B}|$ for the
distribution of gas density in the ICM. This is done by multiplying
the 3D cube obtained in the previous step for a function of the gas
density, $F(n)$.  Based on the results of \citet{Bonafede10}, we will
assume in this work $F(n) \propto n^{0.5}$, which means that for an isothermal
ICM the magnetic field energy density scales with the gas thermal
energy, $B^2 \propto n k_{\rm B} T$.  
We finally normalise our 3D
distribution requiring that the average magnetic field inside a given
radius (which is usually fixed to be cluster core radius) is $B_{0}$.
The constant of normalisation is hence defined as:
\begin{equation}
\beta= \frac{B_{\rm 0}}{\frac{\displaystyle\sum\limits_{r< r_c} |\bf B_i|}{N_{r<r_c}}},
\end{equation} 
where $N_{r<r_c}$ indicates then number of cells within $r_c$
So that $\langle B(r_c) \rangle = B_0$.
This ensures that, on average, the magnetic field follows the
profile: 
\begin{equation}
B(r)=\beta B_0  \left( n/n_{\rm 0} \right)^{0.5},
\end{equation}
where $n_{\rm 0}$ is
the average core gas density.  
We note that the normalisation is slightly  different from the one used in \citet{Bonafede10}, but this
is not influent in the SW sector that we are analysing in this work.
We have chosen this normalisation because this requires to compute only a constant factor, and not 
different scaling factor at different radii, which makes the following computations faster.
Since we are interested in
simulating the RM, only one component of the vectorial B field is of
interest from now on.  We define $B_z$ as the component of the field
along the line of sight. Following the usual convention, it is defined
to be positive when the magnetic field is directed from the source
towards the observed and negative otherwise.\\ $\bullet$ {\bf
  Generation of the Mock RM maps.} Once the magnetic field cube and
the gas density profile are obtained, the RM mock images can be
obtained by integrating Eq. \ref{eq:RM} numerically. The RM images are
then extracted at the position of the observed sources, with respect to the 
X-ray centre of the cluster, and convolved with a Gaussian function 
resembling the restoring beam of the radio
observations. Finally, each of the extracted mock RM images is blanked
following the shape of the RM observed signal for that given source.

\section{Results for the NGC 4039 group and in the relic region}
\label{sec:simulations}
We are interested in determining the magnetic field structure in the
region of the Coma relic, and in understanding if any significant
magnetic field amplification is required to reproduce the observed
trends of $\langle |\rm{RM}| \rangle$ and $\sigma_{\rm{RM}}$ trends.
In contrast to our previous work \citep{Bonafede10}, 
our simulated RM maps have to cover a large field of view, which
includes both the Coma cluster and the SW region where the Coma relic
is located.\\ According to X-ray and optical
observations of the region
\citep{Colless96,Neumann01,Neumann03},
the group seems to be on its first infall onto the Coma cluster. The same scenario is supported by 
a net displacement between the hot gas and the centre of the
group.\\ Despite the complexity of the scenario,
the gas density in the Coma-NGC 4839 system can be modelled to a first
approximation as a double $\beta$-model. 
The parameters for the
Coma cluster are taken from \citet{1992A&A...259L..31B} and rescaled
to the adopted cosmology \citep[see][for further
details]{Bonafede10}.  Not much is known from the literature about the gas
distribution in the NGC 4839 group. This is likely due to the fact that a $\beta$-model fit
can be regarded  only as a first approximation for the gas distribution.
Nonetheless, \citet{Colless96,Neumann01} have derived that the mass of the group
is $\sim$ 0.1 of the Coma cluster. We have hence rescaled the beta model parameters derived for
the Coma cluster in a self-similar way, to obtain a mass that is 0.1 of the mass of Coma.
The resulting parameters for the  NGC 4839 group are given in Table
\ref{tab:betamodel}. The parameters we are using  should be regarded as
first approximations  for the density distribution in the relic region, rather than
as a precise measurement of the gas density profile. Nonetheless, we will show in Sec. \ref{sec:discussion}
that this choice provide an estimate of the gas density at the relic position which is in very good agreement with 
recent results from {\it Suzaku}, as we show in Fig. \ref{fig:bandn} \citep{2013arXiv1302.2907A,2013arXiv1302.4140S}.\\
 Using this setup, we have investigated (i) whether the magnetic
field inferred from the RM in the innermost regions of Coma can
reproduce the observed RMs along the south-west sector of the
Coma- NGC 4839 system;  (ii) whether a change in the magnetic field
power spectrum can reproduce the observed values; (iii) which is is the best magnetic field model
able to fit the data in the group, regardless of the values inferred from
the results obtained for the Coma cluster; (iv) which inferred values
for $n$ and B are needed in this sector to reproduce the observed RMs.

\begin{table}     
\centering
\begin{tabular}{| c | c c c|}    
\hline
\hline
& $n_0$ & $\beta$ & $r_c$\\
Coma cluster        &  3.44 $\times$ 10$^{-3}$  &  0.75  &  290 kpc \\
NGC 4839 group  &  3.44 $\times$ 10$^{-3}$  &  0.75  &  134 kpc \\
\hline
\hline
\end{tabular}
\caption{Gas parameters for the double-$\beta$ model: the central density $n_0$, the value of $\beta$ and of the core radius, $r_c$ are
listed for the Coma cluster and for the NGC 4839 group}      
\label{tab:betamodel}      
\end{table}

\subsection{Standard model: double $\beta$-model with $B_0=4.7 \; \mu$G and $F(n) \propto n^{0.5}$.}
\label{sec:model1}
\begin{figure*}
\centering
\includegraphics[width=0.47\textwidth]{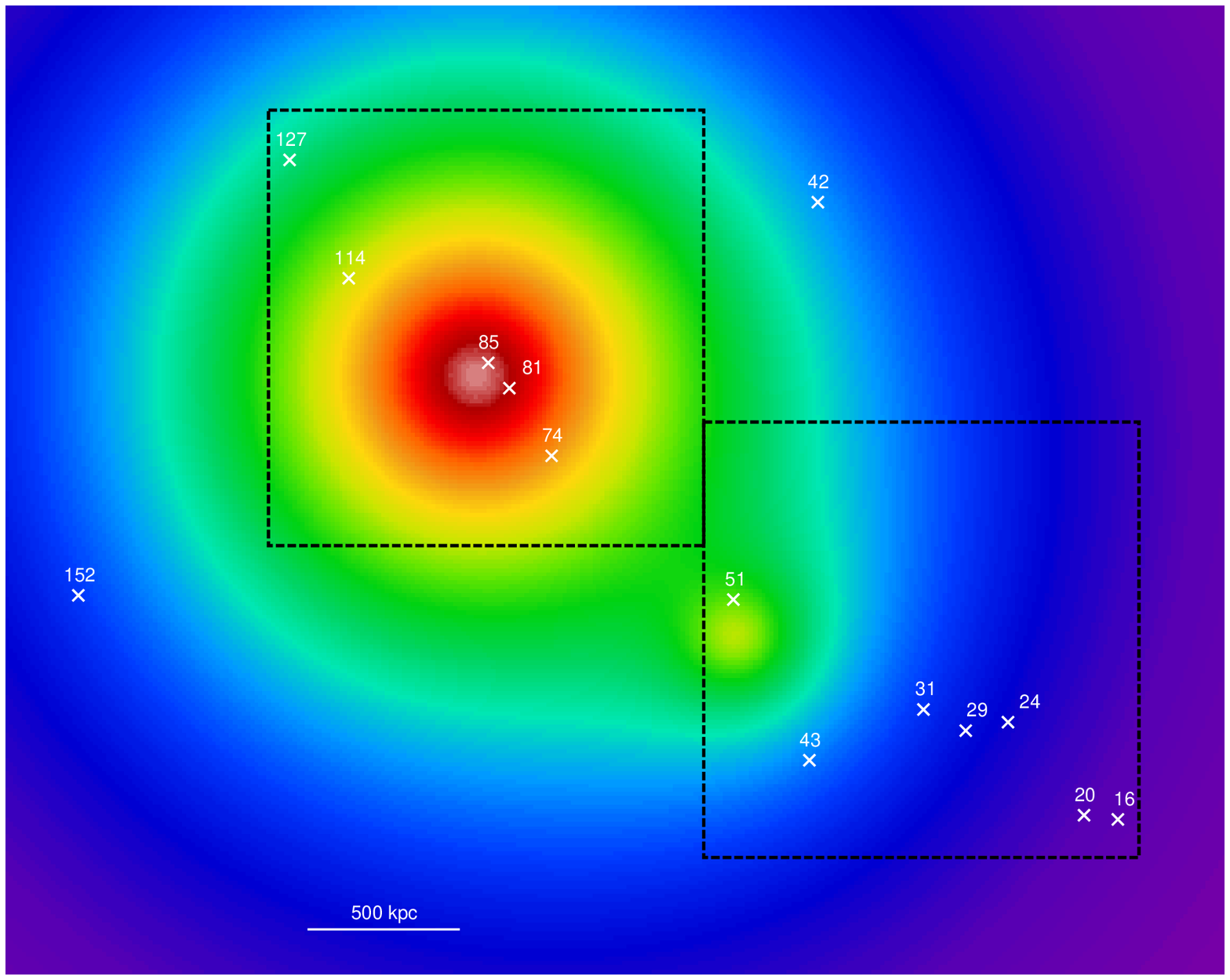}
\includegraphics[width=0.51\textwidth]{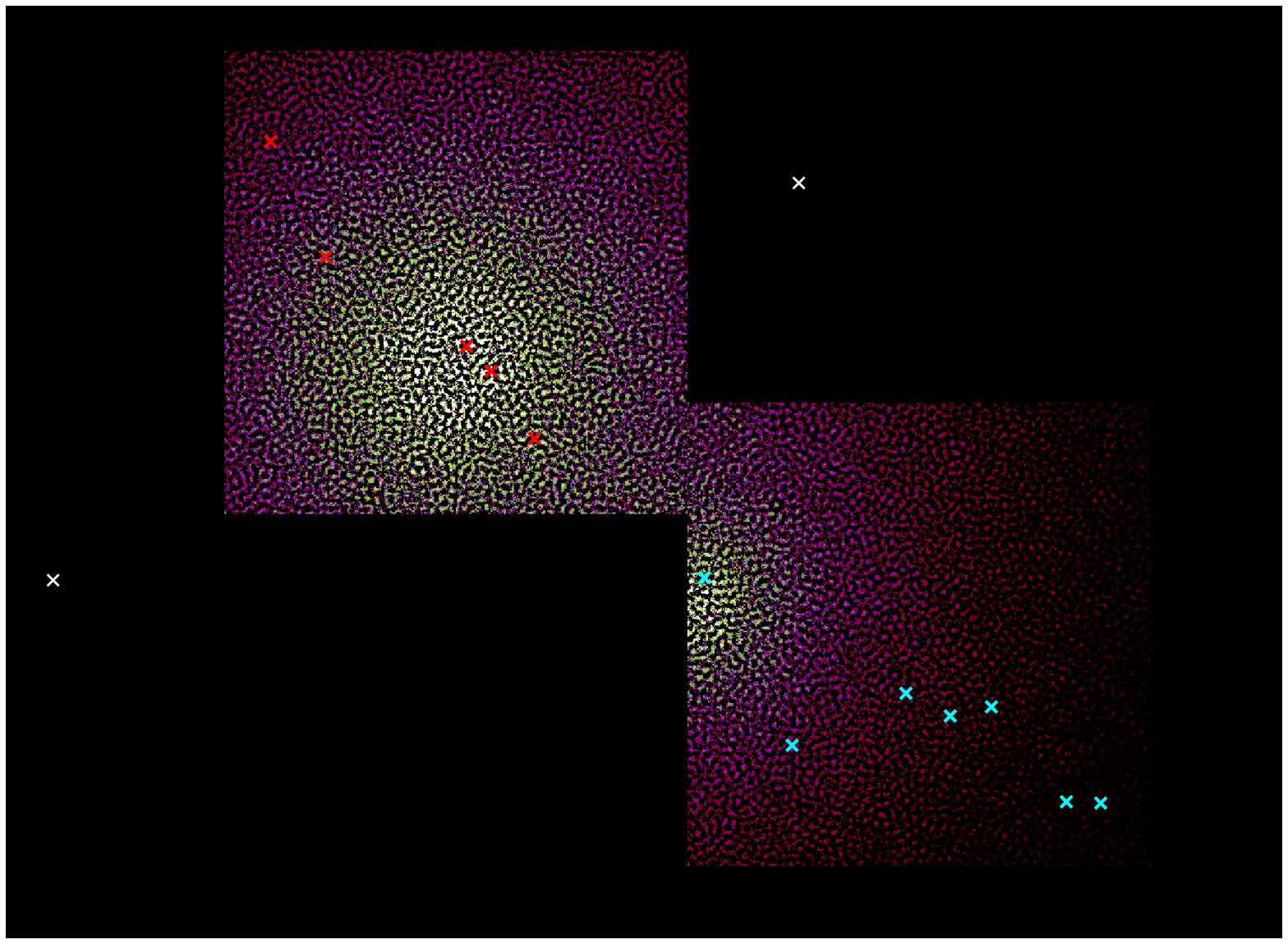}
\caption{{\it Left}: Simulated X-ray emission for the double-$\beta$
  model. Crosses mark the position of the sources in the Coma cluster
  and in the South-West region. Black squares mark the position where
  the RM has been simulated. For display purposes, "5C4. " is omitted in the source labels.
  {\it Right}: Simulated RM maps using the
  parameters of the standard model and two boxes of 1024$^3$ cells (see text). 
  White and red crosses mark the
  position of the sources analysed in Bonafede et al. 2010 (the two
  white crosses label the the position of the sources that are not taken
  into account in this work), cyan
  crosses mark the position of the sources analysed in this work.  }
\label{fig:Xraysim}
\end{figure*}

\begin{figure*}
\centering
\includegraphics[width=\textwidth]{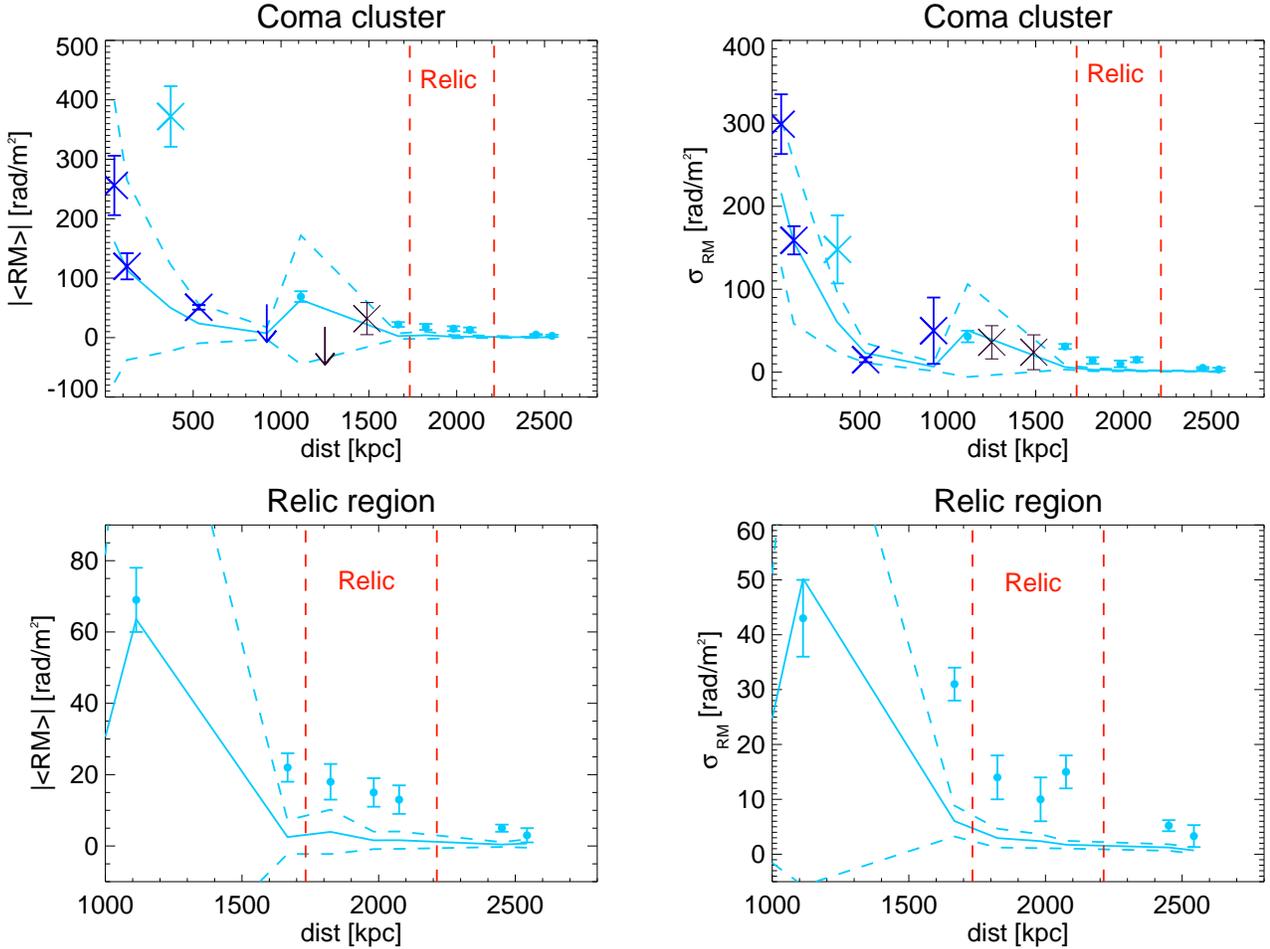}
\caption{$\langle | \rm{RM}| \rangle$ and $\sigma_{\rm{RM}}$ derived from the
  "standard'' model, i.e. considering a double-beta model for the gas
  density and the magnetic field profile with $B_0=$4.7 $\mu$G and
  $F(n)=n^{0.5}$. Solid lines are the mean of the different random
  realisations, dashed lines are the 3$\sigma$ dispersion. {\it Upper
  panels}: crosses are the sources in the Coma cluster from Bonafede et
  al. (2010), dots are the sources analysed here, in the NGC 4839
  group and in the relic region. Cyan symbols refer to the sources in
  the quadrant towards the NGC 4839 group, SW of the Coma cluster,
  black symbols mark the two sources that are out of the simulated
  box. {\it Lower panels}: zoom into the NGC4839 group and into the relic
  region.}
\label{fig:rmsimobs}
\end{figure*}

First, we verified that the new code reproduces within the errors the results found in \citet{Bonafede10}
for the Coma sources.
We have simulated a magnetic field cube assuming a mean magnetic field
of 4.7 $\mu$G within the core radius, and setting the function $F(n)$
to $F(n) \propto n^{0.5}$. The power spectrum is a single power-law,
with scales going from 2 to 34 kpc and a Kolmogorov-like slope.  This combination of parameters
for the 3D magnetic field gives the best reconstruction of the
observed RM trends in the Coma cluster \citep{Bonafede10}. Fig.  \ref{fig:rmsimobs} shows the results obtained with
this new code; they reproduce the results obtained with the {\it Faraday code} within the errors.
However, we note that the different normalisation used here  could in principle produce a larger dispersion
within the cluster core radius.\\
 The radio sources we are
analysing here are located at distances of $\sim$1.1 to 2.3 Mpc from
the cluster's X-ray centre, hence a large field of view needs to be
simulated to include both these new sources and those analysed in
\citet{Bonafede10}.  A single realisation of a 3D box for the
magnetic field, covering the whole field of view
with the resolution of 1 kpc,
 would require a large amount of data ($>$2048$^3$ cells).
However, since the maximum spatial scales are much smaller than
1 Mpc, we can cover the required field of view with two boxes of 1024$^3$ cells each (see Fig. \ref{fig:Xraysim}).
 The magnetic field model has been generated on a
regular mesh of 1024$^3$ cells, with a resolution of 1 kpc.  A second
mesh has been use to cover the relic region, using the same input
model for the magnetic field.  In this second region the magnetic
field is also forced to scale with the square-root of the gas density
(Fig. \ref{fig:Xraysim}). We have performed 20 independent
realisations of the same magnetic field model, and produced mock RM maps. In
Fig. \ref{fig:rmsimobs} the comparison between mock and observed RM trends is
shown. Using $B_0=4.7 \; \mu$G and $F(n) \propto n^{0.5}$, the mean magnetic field within the core radius of the
merging group is $\sim$ 3.6 $\mu$G.
This magnetic field model provides a good description of the
RMs for the Coma cluster sources, both for the $\langle | \rm{ RM}|
\rangle$ and $\sigma_{\rm{RM}}$. It does also provide a good
description for the source 5C4.51, the brightest galaxy of the merging group, 
while it underestimates the RM in the relic region. 
The simple
double-beta model underestimates $\langle |\rm{RM}| \rangle$ and
$\sigma_{\rm{RM}}$ by a factor 6- 8 in the sources observed through the relic
and further out from the NGC 4839 centre.
 This suggests that either the
magnetic field and/or the gas density are enhanced over a region of $\sim 1.5$ Mpc, much
larger than the projected extent of the radio relic. Such departure
from the baseline magnetic field model based on the double
$\beta$-profile for the Coma cluster and the NGC 4839 group indicates
that some additional large-scale mechanism, other than the simple gas
compression provided by the visible structure of the NGC 4839 group,
may be responsible for the observed RM patterns.

\begin{figure}
\centering \includegraphics[width=0.5\textwidth]{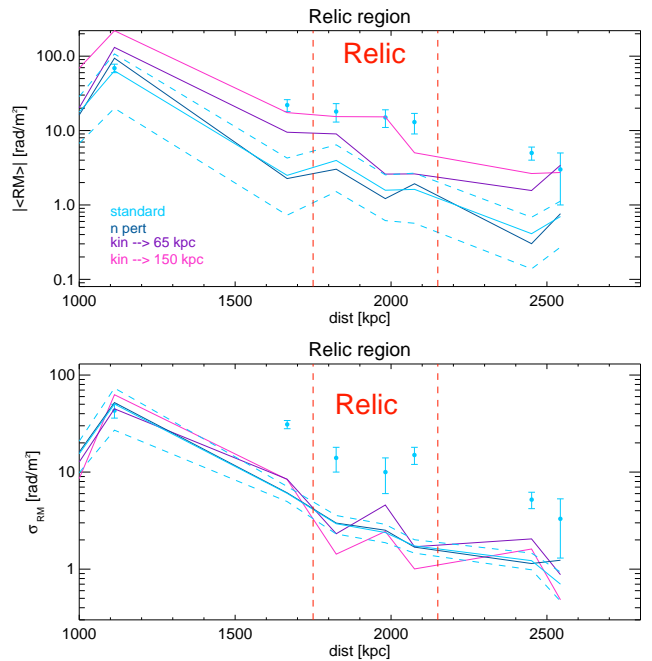}
\caption{$\langle | \rm{RM}| \rangle$ and $\sigma_{\rm{RM}}$ trends from
  magnetic field models with different $k_{in}$, as  shown in the
  plot panel. Cyan solid and dotted lines are like in Fig. \ref{fig:rmsimobs}}
\label{fig:rmPS}
\end{figure}

\begin{table} 
\caption{Magnetic field models}          
\label{tab:Bmodel}      
\centering
\begin{tabular}{|c c c c c|}    
\hline
\hline
Name   &  Gas model &  $k_{\rm in}$ & $k_{\rm out}$ & Gaussian $dn$\\ 
&&&&\\
single $\beta$   &  single $\beta$  & 34 kpc & 2 kpc  & - \\
standard         &  double $\beta$ &  34 kpc & 2 kpc  & -\\
n pert           &  double $\beta$ &  34 kpc  & 2 kpc  & -\\
$k_{\rm in}$ 65      &  double $\beta$ & 65  kpc & 2 kpc  & -\\
$k_{\rm in}$ 150     &  double $\beta$ & 150 kpc & 2 kpc  & $\pm$ 10\%$\times$ n\\
\hline
\hline
\multicolumn{5}{l}{\scriptsize Col. 1: Reference name for the model. Col 2: model for the gas density}\\
\multicolumn{5}{l}{\scriptsize Col. 3 and 4: Corresponding values of $k_{in}$ and $k_{out}$ in the real space.} \\
\multicolumn{5}{l}{\scriptsize Col 5:  Gaussian perturbation for the density model.}\\
\end{tabular}
\end{table}

\subsection{Alternative models for the magnetic field spectrum}
We investigate here whether simple changes to the magnetic field spectrum
in the NGC4839 group and in the relic region can fit the observed RM
values. The parameters adopted for the individual models are listed in
Table \ref{tab:Bmodel}.  The presence of a large-scale infall of gas
along the South-West sectors in the Coma clusters is suggested by
X-ray \citep{Neumann03}, optical and radio \citep{BrownRudnick11}
observations. Bulk and chaotic motions driven by large-scale motions
along this direction may be responsible for the presence of magnetic
field structures on scales larger compared to the innermost regions of
Coma. In order to model this effect, we have simulated a magnetic
field with a power spectrum similar to the one analysed in
Sec. \ref{sec:model1}, but decreasing the value of $k_{\rm in}$. This corresponds 
to assigning more energy into larger eddies, as expected in the case of an ongoing merger event. 
In Fig.  \ref{fig:rmPS} the RM profiles for $k_{\rm in}$ corresponding to
65 kpc and 150 kpc are shown.  An increase in the injection scale up
to a length comparable to the source size has the effect of increasing
the value of $\langle | \rm{RM}| \rangle$ because the biggest modes of the
magnetic field are larger, and contain most of the magnetic field
energy. The effect on the $\sigma_{\rm{RM}}$ depends more on the size
of the source and on how many beams are sampled in the RM images. If
the largest scales in the magnetic field are much larger than the size
of a given radio source, the latter becomes insensitive to large-scale
pattern of RM (which would lead to larger dispersion) because it can
only probe smaller projected scales. The projected size of radio
sources goes from $\sim$10 kpc for 5C4.16 to $\sim$ 120 kpc for 5C4.43 and
the RM signal is not continuous throughout the maps. This causes the
profile of $\sigma_{\rm{RM}}$ not to show a net trend with the
increase of the injection scale. The fit to
$\langle | \rm{RM}| \rangle$ improves as models with larger injection
scales are considered, but such models provide an increasingly poor
fit to $\sigma_{\rm{RM}}$. \\The blue line in Fig.
\ref{fig:rmPS} (``n pert" model) displays the trend of $\langle | \rm{RM}| \rangle$ and
$\sigma_{\rm{RM}}$ for the standard model which fits the Coma central
sources, once Gaussian fluctuation of 10\% in the gas density (and
consequently in magnetic field) profiles are added on scales of $\sim$ 100 kpc.  
These perturbations are consistent with the typical amount of brightness fluctuations 
observed in the centre of  Coma \citep{Churazov12}, and with those measured by
hydrodynamical simulations at such large radii \citep{Vazza13}.
Adding gas perturbations investigates the possible role of enhanced gas clumping along the
South-West sector, which would simultaneously affect the density and
the magnetic (through $\propto n^{0.5}$ scaling) structure. However,
the deviations from the standard profile are small and do not change
significantly the RM statistics.\\ Hence, we can conclude that neither
a change in the power spectrum nor the standard spectrum with the
addition of Gaussian random fluctuations of the order of 10\% are able
to reproduce simultaneously the observed trends of $\sigma_{\rm{RM}}$ and
$\langle | \rm{RM}| \rangle$ in the relic region.

\begin{figure*}
\centering \includegraphics[width=0.8\textwidth]{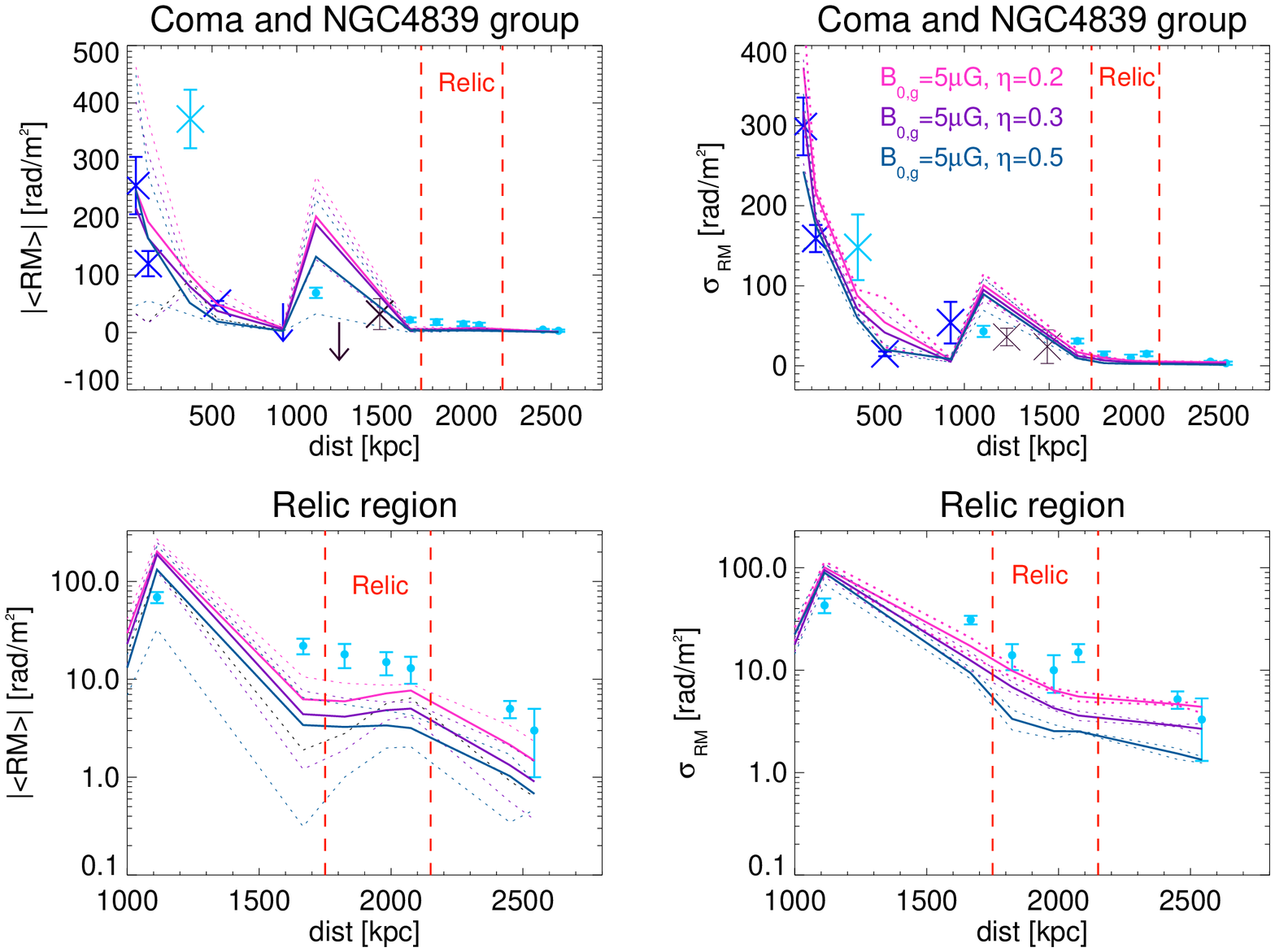}
\centering \includegraphics[width=0.8\textwidth]{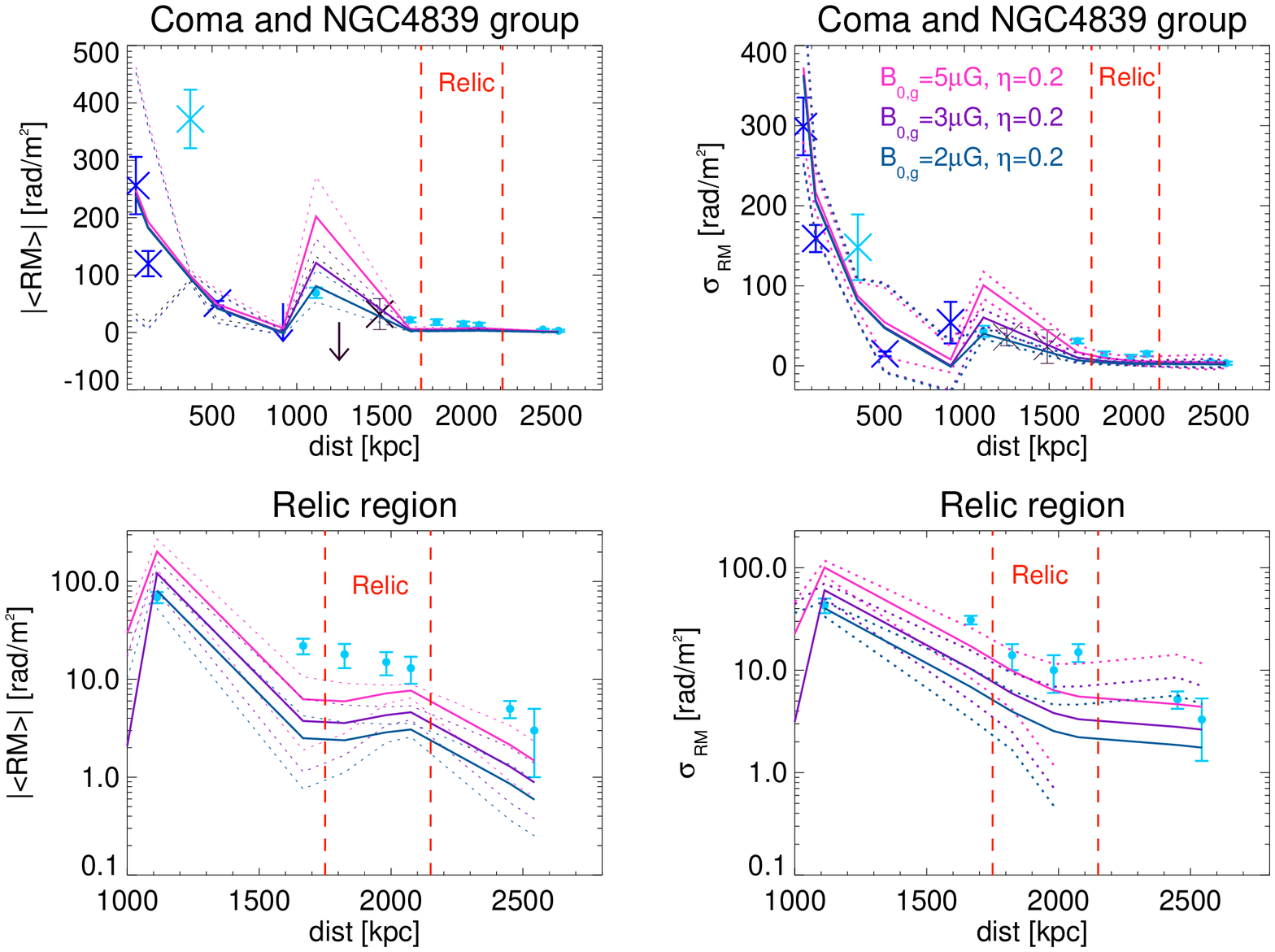}
\caption{$\langle |\rm{RM}| \rangle$ and $\sigma_{\rm{RM}}$ trend versus the
  projected distance from the cluster centre.  Crosses are the points from
  \citet{Bonafede10}, dots are the new points from this paper.  Blue
  symbols mark the sources in the Coma cluster, cyan symbols mark
  those SW from the cluster centre, in the direction of the Coma
  relic. Black crosses mark the position of the sources that are outside the simulated
  box. Red vertical lines indicate the approximate location of the
  Coma relic. Continuous lines display the average among the different realisations, dotted lines
  display the dispersion.
  {\it Top panels}:  $\langle |\rm{RM}| \rangle$ and $\sigma_{\rm{RM}}$  trends are displayed for
  three representative models obtained by changing the value of $\eta_g$, as written in upper right panel. {\it Bottom panels}:
   $\langle |\rm{RM}| \rangle$ and $\sigma_{\rm{RM}}$  trends are displayed for different 
   values of $B_{0,g}$ and $\eta_g=0.2$, as written in the mid-right panel. }
\label{fig:rmgroup}
\end{figure*}

\subsection{The magnetic field in the NGC4839 group}
The group NGC4839 can be modelled as an independent group
falling into the Coma cluster. In Sec. \ref{sec:model1}  we have shown that the scaling
$B(r)\propto n^{0.5}$ inferred from Coma  gives a reasonable description of the RM values
at the centre of the group,  but
does not reproduce within 3$\sigma$ the values of the other sources
analysed in this work. Although the $\beta$-model used for the group  does not provide an accurate
description of the gas distribution, we attempt to fit the RM values for the sources 
in the SW sector by assuming an independent magnetic field model for the NGC4839 group.
We have assumed the following scalings for the magnetic field in the group:
\begin{equation}
 B_{\rm group}(r) =\hat B_{g} \left( \frac{n_g(r)}{n_0} \right)^{\eta_g}
\end{equation}
and for the Coma cluster:
\begin{equation}
B_{\rm Coma}(r)  =\hat{B}\left( \frac{n_c(r)}{n_0} \right)^{\eta}
\end{equation}
and the same values of $k_{\rm in}$ and $k_{\rm out}$ as found in \citet{Bonafede10}.
We have realised independent grids of 3D magnetic fields for the two systems, 
and added them together in the regions of interest. The fact that $\nabla \cdot {\bf B} =0 $
by construction in both grids allows us to perform this operation  preserving the zero-divergence condition
with the same accuracy.  The centre of the group along the line of sight is placed on the same plane as the Coma cluster's centre.
We have then obtained mock RM images by integrating numerically the equation
\begin{equation}
RM=\int{(B_{\rm{Coma},z}\times n_c+B_{\rm{group},z}\times n_g) dl}
\end{equation}
changing the values of $B_{0,g}$ from 2 to 10 $\mu$G and $\eta_g$  from 0.2 to 0.5.
Since flat magnetic field profiles for the group could affect also the RM values of the Coma cluster sources,
we have compared mock and real RM images for all the sources in the field. In Fig. \ref{fig:rmgroup} we show 
the results for the most representative cases among those analysed. In the top panels, three models with different $\eta_g$  are shown. 
We note that as the profile flattens (i.e. $\eta_g$ decreases), the RM values at the centre of the group
increase, as expected since the RM is an integrated quantity along the line of sight, and it is sensitive to
the magnetic field in the outer parts of the group. Hence, although  using flatter profiles 
$\sigma_{\rm{RM}}$ and $\langle |\rm RM| \rangle$  become closer to the observed values in the relic region, they provide a worse
fit in the centre of the group. Even a flat magnetic field profile ($\eta_g=0.2$) is not able to reproduce the observed 
values of $\sigma_{\rm{RM}}$ and $\langle |\rm RM| \rangle$ irrespective of  $B_{0,g}$. As shown in the bottom panels
of Fig. \ref{fig:rmgroup},  values of $B_{0,g}$ of the order of 2 $\mu$G are required for the centre of the group, while higher values would be needed
in the relic region and beyond.\\
We conclude that, in the double $\beta$-model approximation, there is no obvious combination of values for $B_{0,g}$ and $\eta_g$ which is able 
to reproduce the observed RMs in the SW sector of the Coma cluster.  Even flat profiles
with $\eta_g=$0.2 fail in describing the trends of $\sigma_{\rm{RM}}$ and $\langle |\rm RM| \rangle$ in the SW region.\\

\section{Discussion}
\label{sec:discussion}

We tried to model the 3D structure of the magnetic field in the Coma
cluster and along the relic sector by simulating different configurations
of ${\bf B}$, similar to what
has been already done in the literature \citep[e.g.][]{Murgia04,Bonafede10,Vacca12}.\\
The extrapolation of the magnetic field model that successfully reproduces the trend of RM
in the  Coma cluster  \citep{Bonafede10} does not reproduce the
trend of RM along the SW sector. The values of 
$\langle |\rm RM| \rangle$ and $\sigma_{\rm RM}$ are nearly one order of magnitude lower 
than the observed values.\\
Several independent works have found large-scale accretion patterns in the SW region
of the Coma cluster,  using X-ray observations \citep[][see however \citealt{Bowyer04} for a different interpretation ]{Neumann03,OgreanComa,2013arXiv1302.2907A,2013arXiv1302.4140S} , optical observations \citep{Colless96,Neumann01, BrownRudnick11},
and radio observations \citep[e. g.][]{BrownRudnick11}. 
Therefore, a significant departure
from the simple $\beta$-model profile along this direction is very likely.\\
We tested this scenario simulating a double-$\beta$-model, which
considers a second spherical gas concentration coincident with the group NGC4839. 
We assumed for the group a core radius and a central density which are
the rescaled version of the Coma cluster for a system with one tenth of the mass, as indicated by  \citet{Neumann01}.
We have produced 3D magnetic field simulations for this double $\beta$-model, normalising the
mean magnetic field at 4.7 $\mu$G within the Coma core radius, and scaling the magnetic field profile
with the square-root of the gas density throughout the whole simulated volume. This choice gives a mean magnetic field within the core radius of the group
of 3.6 $\mu$G.
We find that at 99\% confidence level, the values of  $\langle |\rm{RM}| \rangle$ and $\sigma_{\rm RM}$ at the location of 5C4.51, the brightest galaxy
of the  group, can be explained by the double $\beta$-model. However, along a sector of $\sim 1 \rm Mpc$ further to the SW, the observed values of the Faraday Rotation are still larger than
the simulated double $\beta$-model by a factor $\sim 6$.
Basic modifications to the 3D model that we have tried (e.g. by imposing different spectra, 
or adding some amount of gas clumping to the simulated gas model) are 
unable to simultaneously reproduce $\langle |\rm{RM}| \rangle$ and $\sigma_{\rm RM}$
for all the sources. 
We have also attempted to reproduce the RM values in the group by fitting independently a 
magnetic field model for the group and then adding the contribution of the Coma cluster. 
Even very flat magnetic field profiles  ($\eta_g=$0.2) are unable to reproduce the trends of $\langle |\rm RM| \rangle$ and 
$\sigma_{\rm RM}$ along the entire SW sector out to the relic and beyond.
We note that magnetic field profiles flatter than $\eta=0.2$ would be in disagreement with previous works \citep[e.g.][]{2006A&A...460..425G,2008A&A...483..699G,Bonafede10}
and also with cosmological MHD simulations \citep[e.g.][and ref. therein]{Donnert10,sk13,2008SSRv..134..311D}.\\
Hence, we conclude that the RM data along the SW sector of the Coma cluster
require additional amplification of the magnetic field. 
\subsection{Limits to the magnetic field and gas density in the relic region}
\begin{figure}
\centering \includegraphics[width=0.5\textwidth]{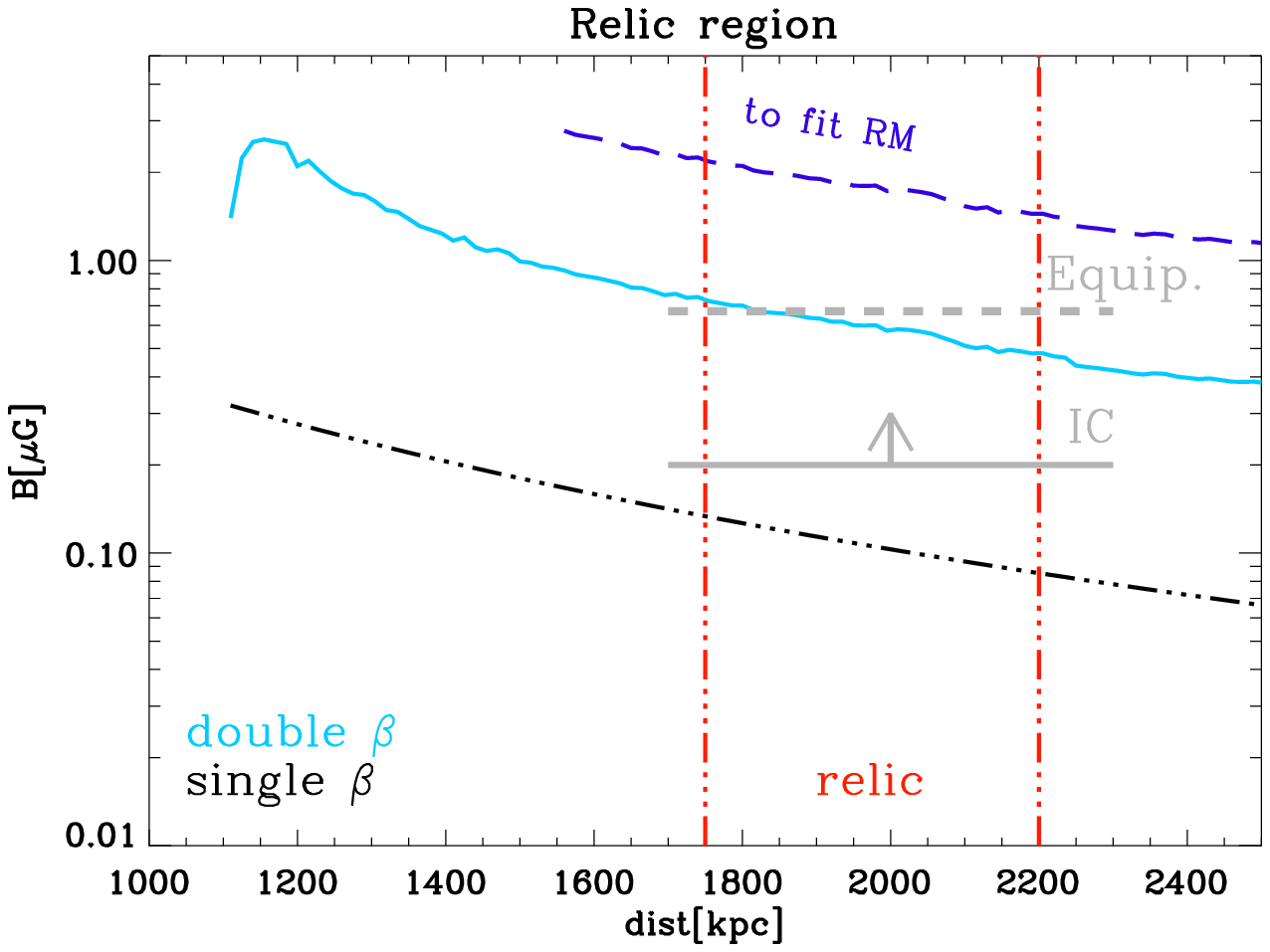}
\centering \includegraphics[width=0.5\textwidth]{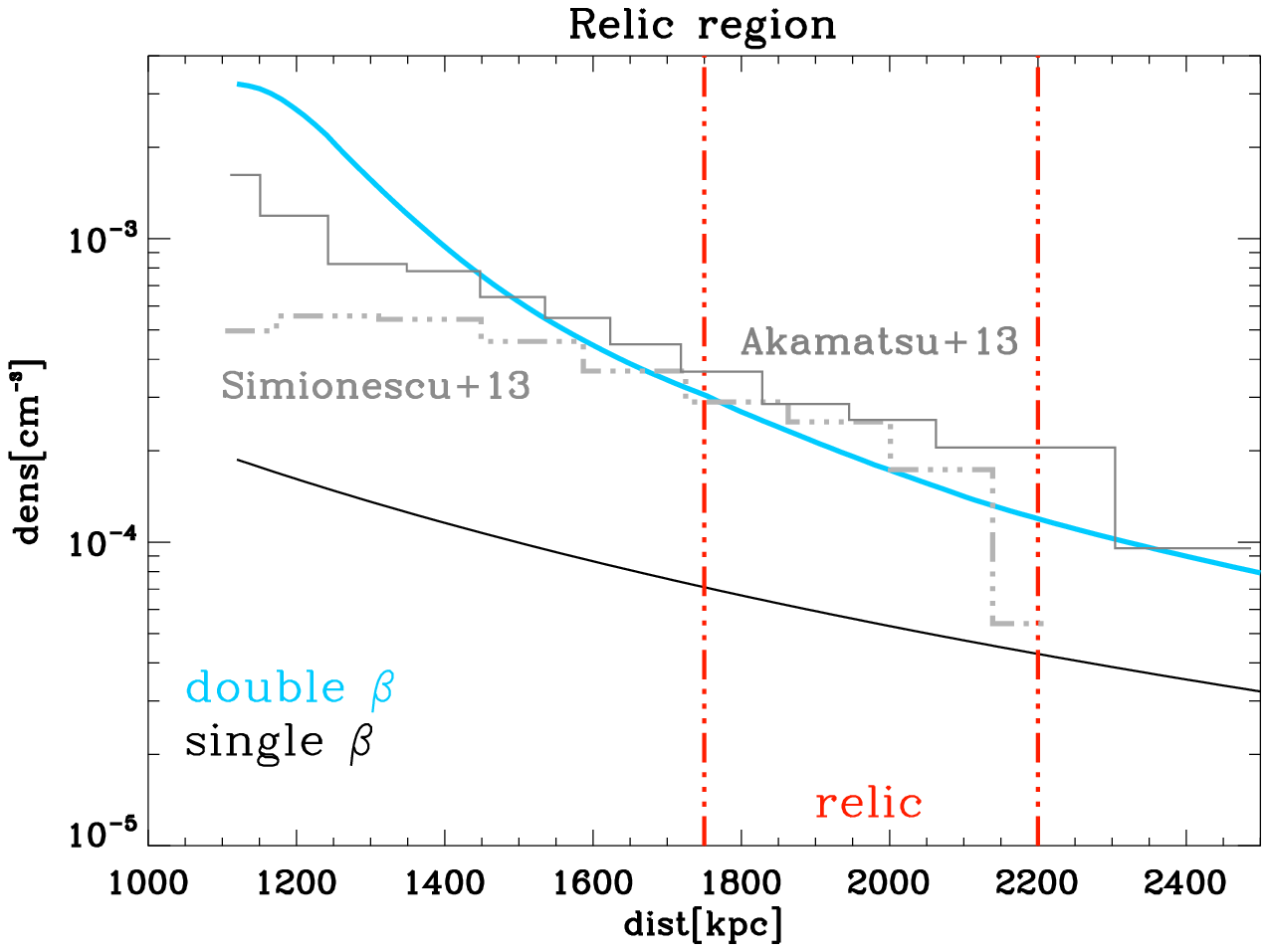}
\caption{Top: Magnetic field profiles in the group and relic region. The black dashed-double dotted line displays the profile
  obtained by Bonafede et al (2010) for the Coma sources, the cyan
  line  displays the profiles from the standard model, assuming a double-$\beta$ model for the gas density, $B_0=4.7 \; \mu$G and $F(n)=n^{0.5}$.
  The blu dashed line displays the magnetic field needed
  to fit the RM data in the relic region and beyond. Overplotted are the lower limits from the Inverse Compton
  \citep{2006A&A...450L..21F},  and the estimate from Equipartition \citep{2003A&A...397...53T}. 
  Bottom: Gas density profile in the group and relic region. The black line displays the single $\beta$-model for Coma, 
  the cyan line displays the double $\beta$-model for Coma and NGC4839 group. Grey lines display the profile derived by Suzaku observations.}
    \label{fig:bandn}
\end{figure}

In order to understand how the gas and the magnetic field contribute to the RM enhancement, it would be necessary to derive independent estimates
or limits for the two, separately. Although it is not possible to derive limits for the magnetic field, recent
{\it Suzaku} X-ray observations provide useful constraints on the gas density at the position of the relic \citep{2013arXiv1302.2907A,2013arXiv1302.4140S}.
 From the brightness profiles present in the two recent papers, we have computed
the corresponding density profiles assuming a constant temperature. The two profiles are shown in
 Fig. \ref{fig:bandn}. The small differences between the two are due to 
  the different regions used for spectral analysis, and to slightly different de-projection assumptions
 used in the two works. 
Although the double $\beta$-model that we are using is only a first-order approximation
for the real gas density distribution in the system, the density profile at the position of the relic is compatible with
the real gas density within a$\sim50$ percent (see Fig.\ref{fig:bandn}), while it is higher by a factor $\sim$ 5 
inside the core radius of the group.
This suggests that the larger observed values of Faraday Rotation along the SW sector cannot be explained by 
a medium much denser than what assumed here, unless it is too cold ($T<10^{7} \rm K$) to emit in X-rays. 
However, the overall good agreement between the outer gas density profiles obtained through X-ray and  SZ observations
 \citep{PlanckComa, 2013ApJ...763L...3F} makes this scenario unlikely.\\
A uniform boost in the magnetic field by a factor $\sim 6$ would reconcile with the data in a rather simple way, yielding an average level of magnetic field of $\sim 3-4 \; \mu \rm G$ at the location of the relic.
 This value would be about 3 times higher than
the radio equipartition values \citep{2003A&A...397...53T}, and compatible with the lower limits provided by the Inverse Compton analysis \citep[e.g.][]{2006A&A...450L..21F}.
The boost factor can be reduced to $\sim 3$  if the gas density suggested by the most recent X-ray observations is
used, yielding a value of $\sim 1.5-2 \; \mu \rm G$ at the location of the relic.
Magnetic field with strength of the order of few $\mu$G in radio relics are found by cosmological simulations \citep{sk08}
and using radio equipartition estimates \citep[e.g.][]{2009A&A...494..429B, 2010A&A...511L...5G}.
However, our data suggest that the required magnetic boost  is not limited to the relic region - which is $\sim$ 400 kpc wide- but extends 
up to a distance of $\sim  1.5 \;  \rm Mpc$ from the group centre, although the large uncertainties in the region beyond the relic do not allow
significant constraints on the gas density (see Fig. \ref{fig:bandn}).\\
\subsection{Implications on dynamics}
The derived values of the magnetic field along the SW sector have no important  impact on the cluster dynamics.
The plasma beta, $\beta_{\rm pl} $, can be estimated as
$\beta_{\rm pl} \approx 2 c^2_{\rm s}/v^2_{\rm A}$, where $c_{\rm s}$ is the gas sound speed and $v_{\rm A}$ is the Alfv\'{e}n speed.
The presence of the X-ray emitting gas yields  $\beta_{\rm pl} \approx10-100$, and therefore the magnetic pressure is still dynamically irrelevant.\\
A rather natural way of producing such large-scale magnetic fields is a cosmic filaments along the SW direction, a possibility that has
been already suggested elsewhere \citep{Finoguenov03,BrownRudnick11}, and that emerges from MHD cosmological 
simulations as well \citep{2005ApJ...631L..21B,2005JCAP...01..009D}.
However, the observed values of $\sigma_{\rm RM}$ indicate a broad spectrum of fluctuations, in the range 
2 - 50 kpc at least, which are not expected in cosmic filaments. Instead, cosmological simulations indicate
that  the topology of the magnetic field  
in filaments is ordered and uniform, because the eddy turnover-time of turbulent motions is larger than the filaments' age,
and turbulence is not fully developed \citep[e.g.][]{Ryu08}.
The observed dispersion in  the RM for the sources observed through the relic indicates  that turbulent motions
are present in the relic region.  \\
If a shock is present across the Coma relic, as suggested by recent 
X-ray observations \citep{OgreanComa,2013arXiv1302.2907A},  some magnetic field amplification is expected.
It was recently shown by \citet{2012MNRAS.423.2781I} that most of the amplification is due to the compression
of the ICM plasma, while turbulence should play a minor role, although under certain circumstances vorticity generated by compressive and baroclinic effects
 across the shock discontinuity can lead to a sufficient amplification of the magnetic field.
We note that  the observed RMs require an enhancement of the
$B_{//} \times n$ quantity not only in the region of the relic, but over the full SW sector ($\sim$ 1.5 Mpc).
In agreement with this result is the presence of a weak radio emission connecting the halo, the relic, and the radio galaxy 5C4.20.
 Indeed,
the radio halo and the radio relic of the Coma cluster are connected through a low-brightness radio bridge
\citep{1989A&A...213...49V,1993ApJ...406..399G,BrownRudnick11}, and on the Western side of the relic a similar low-brightness
radio bridge connects the relic with the radiogalaxy 5C4.20 \citep{Giovannini91}. 
In addition, single dish observations of the Coma cluster have shown a large amount of diffuse synchrotron  emission in the SW region of the cluster, extending to at least $\sim$ 2.4 Mpc west of the centre
of the Coma cluster's radio halo \citep{Kronberg07,BrownRudnick11}.
Hence,  the observed presence of large-scale relativistic electrons and magnetic fields, probed by radio observations, 
 the elongated X-ray and SZ  morpohologies  \citep{Neumann01,Finoguenov03,PlanckComa},
  and the hint of a weak shock, $M \sim 2$, at the relic location \citep{OgreanComa,2013arXiv1302.2907A}  suggest that a more inhomogeneous 
  large-scale accretion event is taking place. In this framework, the  virialisation of the kinetic energy would not be completed yet,  yielding to a
  magnetic field and gas density amplification.  
 Although in this work we do not attempt to model further the dynamics of the
ICM along the SW direction, the agreement between the density profile used here and the one derived by recent {\it Suzaku} observations indicate that the
limits on the magnetic field at the position of the relic are robust.\\
A more sophisticated analysis can be done by using more realistic distributions for the ICM gas in the SW sector, and
relaxing the assumption of an isotropic magnetic field.  A possibility to explain the observed RM trends would be a large-scale component of the magnetic field, aligned 
with the filament which is accreting matter into Coma, plus smaller scale components that would give the observed RM dispersion.
Such a configuration, opportunely projected along the line of sight, could in principle produce the observed $\sigma_{\rm RM}$ and $\langle |\rm RM| \rangle$ trends.
 We will tackle this issue in a forthcoming work.

\section{Conclusions}
\label{sec:conclusion}
In this work we have presented new RM data for seven sources located across
the Coma relic. The trends of $\sigma_{\rm RM}$ and $\langle |\rm RM| \rangle$
have been analysed together with the data presented in \citet{Bonafede10}, to  probe 
the magnetic field properties in the outskirts of the cluster, where the radio relic is located.
We have presented a new tool to interpret the RM data by comparing mock and real RM observations.
The main results can be summarised as follows:
\begin{itemize}
\item{Both $\sigma_{\rm RM}$ and $\langle |\rm RM| \rangle$ decrease going from the centre of the NGC4839 group towards the
outskirts of the cluster, indicating that both the gas density and the magnetic field profile decrease radially. No evident jump is found at the position of the relic.}\\
\item{The observed values of $\sigma_{\rm RM}$ indicates the presence of a magnetic field which fluctuates over a range of spatial scales, possibly
indicating turbulent motions in the SW region of the cluster, across the relic and beyond.}\\
\item{Both $\sigma_{\rm RM}$ and $\langle |\rm RM| \rangle$  are higher than predicted by simply extrapolating the magnetic field and the gas density
profiles which, instead,  give the best fit for the RMs in the Coma cluster. }\\
\item{When a double $\beta$-model is used to describe the gas density profile in the system Coma - NGC4839 group, the best fit values found for the
magnetic field in the Coma cluster, e.g. $B_0=4.7 \,\mu$G and $F(n) \propto 0.5$ \citep{Bonafede10} reproduce the values of
$\sigma_{RM}$ and $\langle |\rm RM| \rangle$ only for the brightest source of the group: 5C4.51 (NGC4839). 
This magnetic field model gives an average magnetic field of 3.6 $\mu$G within the group core radius.
The  trends of $\sigma_{\rm RM}$ and $\langle |\rm RM| \rangle$
for the remaining six sources in the SW sector are  underestimated by a factor 6 - 8. By comparing these data with recent gas density estimates from {\it Suzaku} \citep{2013arXiv1302.2907A,2013arXiv1302.4140S} 
we have derived that a boost of the magnetic field by a factor $\sim$ 3 is required.}\\
\item{The magnetic field amplification does not appear to be limited to the relic region, but it must occur throughout the whole SW sector ($\sim 1.5$ Mpc).}\\
\item{There is no simple combination of $\beta$-models for Coma and the NGC4839 group  that can
reproduce the observed trends of $\sigma_{\rm RM}$ and $\langle |\rm RM| \rangle$  at the same time. Similarly, neither a change in the magnetic field power spectrum in the SW sector, neither
adding gaussian perturbation to the gas density and to the magnetic field  can help is reconciling observed and simulated values.}\\
\item{The analysis of these data together with the available results from X-ray \citep{Neumann03,OgreanComa,2013arXiv1302.2907A,2013arXiv1302.4140S}, optical \citep{Colless96,Neumann01, BrownRudnick11}, SZ \citep{PlanckComa},  and radio \citep{Giovannini91,Kronberg07,BrownRudnick11} observations indicate that the most plausible scenario is the one in which a  large-scale accretion event, 
followed by a yet incomplete virialisation of kinetic energy,  is taking place in the SW region of the Coma cluster, yielding to a
  significant magnetic field amplification}.  \\
  \item{In order to reconcile mock and observed RM trends in the SW region, the magnetic field must have been amplified by a factor $\sim$3. 
  In this case, the magnetic field across the relic region should be then $\sim$ 2 $\mu$G, which is consistent with IC upper limits.} \\
  \end{itemize}
  
  \bigskip
{\bf Acknowledgments} The authors thank K.Dolag for fruitful discussions and for the use of  {\it Pacerman}.
AB, MB, and FV acknowledge support by the
research group FOR 1254 funded by the Deutsche Forschungsgemeinschaft:
``Magnetization of interstellar and intergalactic media: the prospect
of low frequency radio observations''.
 This research has made use of the NASA/IPAC Extragalactic
Data Base (NED) which is operated by the JPL, California institute of
Technology, under contract with the National Aeronautics and Space
Administration.  
\bibliographystyle{aa}
\bibliography{master}

\end{document}